\documentclass[twocolumn,english,aps,prb,floatfix,amssymb,superscriptgroupedaddress]{revtex4}
\usepackage[T1]{fontenc}
\usepackage[latin9]{inputenc}
\usepackage{color}
\usepackage{array}
\usepackage{rotating}
\usepackage{float}
\usepackage{multirow}
\usepackage{amsmath}
\usepackage{amssymb}
\usepackage{graphicx}
\usepackage{esint}

\makeatletter

\providecommand{\tabularnewline}{\\}

\@ifundefined{textcolor}{}
{%
 \definecolor{BLACK}{gray}{0}
 \definecolor{WHITE}{gray}{1}
 \definecolor{RED}{rgb}{1,0,0}
 \definecolor{GREEN}{rgb}{0,1,0}
 \definecolor{BLUE}{rgb}{0,0,1}
 \definecolor{CYAN}{cmyk}{1,0,0,0}
 \definecolor{MAGENTA}{cmyk}{0,1,0,0}
 \definecolor{YELLOW}{cmyk}{0,0,1,0}
}

\usepackage{babel}
\usepackage[bookmarks=false,linkcolor=blue,urlcolor=blue,colorlinks,citecolor=blue]{hyperref}
\makeatother

\usepackage{babel}
\begin{document}

\title{Topological Superconducting Phases of Weakly Coupled Quantum Wires}

\author{Inbar Seroussi}

\author{Erez Berg}

\author{Yuval Oreg}

\affiliation{Department of Condensed Matter Physics, Weizmann Institute of Science,
Rehovot, Israel 76100}
\begin{abstract}
An array of quantum wires is a natural starting point in realizing
two-dimensional topological phases. We study a system of weakly coupled
quantum wires with Rashba spin-orbit coupling, proximity coupled to
a conventional s-wave superconductor. A variety of topological phases
are found in this model. These phases are characterized by ``Strong''
and ``Weak'' topological invariants, that capture the appearance of
mid-gap Majorana modes (either chiral or non-chiral) on edges
along and perpendicular to the wires. In particular, a phase with
a single chiral Majorana edge mode (analogous to a $p+ip$ superconductor)
can be realized. At special values of the magnetic field and chemical
potential, this edge mode is almost completely localized at the outmost
wires. In addition, a phase with two co-propagating chiral edge modes
is observed. We also consider ways to distinguish experimentally between
the different phases in tunneling experiments.\textcolor{red}{}
\end{abstract}

\date{\today}

\maketitle

\section{Introduction}

Topological insulators and superconductors have received much attention
in the past few years\cite{hasan2010colloquium,qi2011topological,bernevig2013topological}.
Such phases are characterized by a gap for bulk excitations, while
the boundaries support topologically protected gapless edge states.
In addition, topological defects in these phases may carry exotic
zero energy excitations with unusual properties. For instance, defects
in topological superconductors (such as vortices in two-dimensional
chiral p-wave superconductors\cite{read2000paired} or edges of one-dimensional
spinless p-wave wires \cite{kitaev2001unpaired}) support localized
states known as Majorana zero modes. These zero modes have non Abelian
properties, and have been proposed as possible ingredients for a topological
quantum computer\cite{Kitaev20032}.

Currently, the most promising experimental proposal for realizing
Majorana zero modes in solid state devices involves quasi-1D semiconductor
nano wires with strong spin orbit coupling, such as InAs or InSb, proximity
coupled to a s-wave superconductor \cite{lutchyn2010majorana,oreg2010helical}.
The main advantage of this proposal is its simplicity: it does not
require any exotic materials, but rather involves only conventional
semiconductors and superconductors. Recent experiments have detected
signatures of Majorana zero modes in heterostructures of semiconducting
quantum wires and superconductors\cite{mourik2012signatures,das2012zero,deng2012anomalous,churchill2013superconductor,rokhinson2012fractional}.

In a two dimensional system of a spinless p-wave superconductor with
pairing potential $\Delta(\mathbf{k})\sim\mathbf{k}$ a chiral p-wave
with a gapless edge state can be formed\cite{read2000paired}. Other
possible realizations of this phase are presented in Refs. \onlinecite{alicea2010majorana, sau2010generic, fu2008superconducting}.
A wider variety of phases is studied in Ref. \onlinecite{asahi2012topological}
using a toy model of spinless electrons in a two dimensional p-wave
superconductors.

\textcolor{black}{The topological phases can be classified based on
the symmetries (i.e., time-reversal, particle-hole, and chiral symmetries),
and the dimensionality of the system. The topological classification
is summarized in the ``periodic table'' studied in Refs. \onlinecite{kitaev2009periodic}
and,  \onlinecite{schnyder2008classification}. The spinless p-wave
superconductor is in class $D$ (particle-hole symmetric, but not
time reversal symmetric), and is characterized by a $\mathbb{Z}$
invariant counting the number of gapless chiral Majorana modes on
the boundary of the system. In the presence of translational symmetry, one can also define two $\mathbb{Z}_{2}$ invariants which count the
parity of the number of boundary Majorana modes in each direction.}\textcolor{red}{{}
}\textcolor{black}{The $\mathbb{Z}$ number is referred to as a }\textit{\textcolor{black}{strong
index,}}\textcolor{black}{{} and the $\mathbb{\mathbb{Z}}_{2}$ numbers
are }\textit{\textcolor{black}{weak indices}}\textcolor{black}{\cite{fu2007topological}.
Here, we will be interested in identifying these phases in a physically
realizable model.}

\begin{figure}
\begin{centering}
\includegraphics[bb=0bp 0bp 496bp 287bp,scale=0.4]{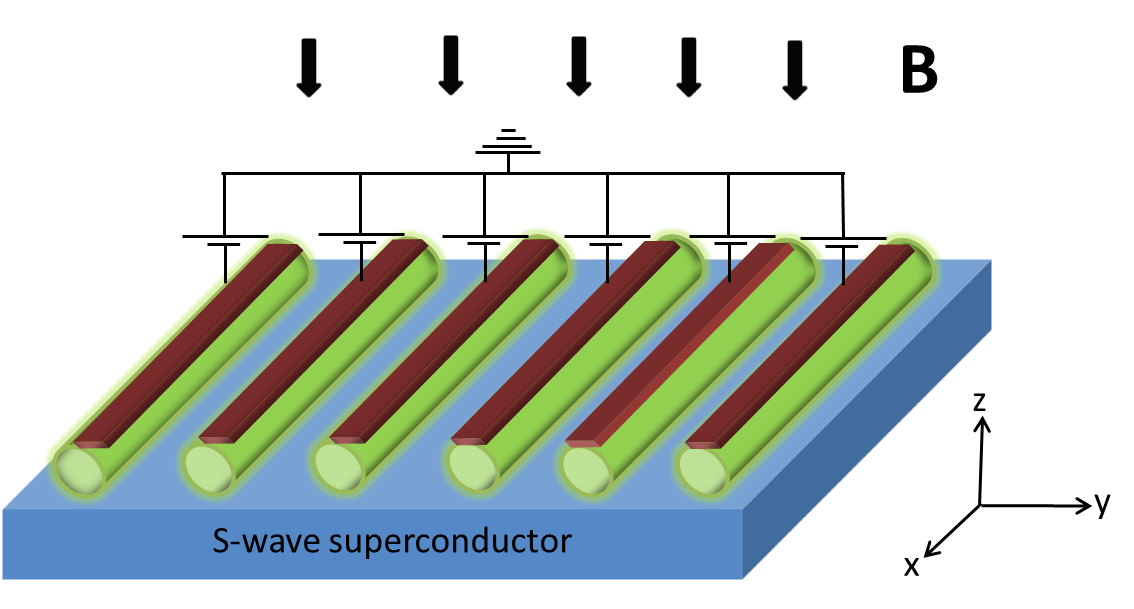}
\par\end{centering}

\caption{Schematic view of the system. An array of semiconducting quantum wires
proximity coupled to an s-wave superconductor. The density of electrons
in the wires is controlled by nearby metallic gates. A magnetic field
is applied perpendicular to the wires.\label{fig:A-schematic-picture of the SF-1} }
\end{figure}

In this work, we demonstrate that quantum wires (or ribbons) can be
used as a platform to realize a rich variety of topological superconducting
phases. An array of weakly coupled wires, such as the one discussed
above, is a natural starting point to realize a two-dimensional phase,
analogous to the chiral p-wave phase of Read and Green\cite{read2000paired},
which supports chiral Majorana modes in its boundaries. A graphical
illustration is presented in Fig. \ref{fig:A-schematic-picture of the SF-1}.
We study how varying experimentally controllable parameters, such
as the magnetic field and chemical potential, allows to tune into
these phases.

In particular, we show that there is a choice of parameters such that
the counter-propagating chiral edge states are almost completely localized
on the two outmost wires, allowing the observation of the chiral
phase even in an array with only a few wires. One also finds a phase
with two co-propagating chiral modes localized at each edge. In the
phase with a chiral edge mode, an orbital magnetic field perpendicular
to the plane of the wires induces vortices which carry Majorana zero
modes at their cores. We also show how the zero energy density of
states (DOS) changes as a function of the orbital field. We discuss
experimental signatures that can be used to identify these phases,
through scanning tunneling microscopy into the outmost wires.

The paper is organized as follows. In Sec. \ref{sec: Overview  Topological  supercondu},
we briefly review the topological superconducting phases that can
arise in a two-dimensional system with translational symmetry, through
the model of spinless fermions that was introduced in Ref. \onlinecite{asahi2012topological}.
We then consider an array of weakly coupled semiconducting wires
of the type studied in Refs. \onlinecite{lutchyn2010majorana,oreg2010helical}
In Sec. \ref{sec:spinfull model}. We explore the phase diagram of
the system as a function of experimentally controllable parameters,
and the structure of the edge states in the chiral phases. In Sec.
\ref{sec:The orbital effect} we consider the effect of an orbital magnetic
field. In Sec. \ref{sec:Experimental-signature} we study the experimental
signatures of the different phases. Sec. \ref{sec: conclusions} summarizes
our main results and conclusions.

\section{{\normalsize{Overview: Topological superconducting phases of spinless
fermions in a two dimensions}}}

\label{sec: Overview  Topological  supercondu}

In this section, for the purpose of illustration and to set up the
framework, we will review the analysis of a toy model of spinless
fermions hopping on a square lattice with a p-wave pairing potential.
This model was introduced and analyzed in Ref. \onlinecite{asahi2012topological}.
The tight binding Hamiltonian is:

\begin{eqnarray}
\mathcal{H}&=&\sum_{i,j}[-t_{x}\psi_{i,j}^{\dagger}\psi_{i+1,j}-t_{y}\psi_{i,j}^{\dagger}\psi_{i,j+1}\nonumber \\
&-&\mu(\psi_{i,j}^{\dagger}\psi_{i,j}-\frac{1}{2})+d_{x}\psi_{i,j}^{\dagger}\psi_{i+1,j}^{\dagger}\nonumber \\
&+&id_{y}\psi_{i,j}^{\dagger}\psi_{i,j+1}^{\dagger}+h.c.],\label{eq: spinlees Hij}
\end{eqnarray}
when $t_{x}$ and $t_{y}$ are the tunneling matrix elements in the
$x$ and $y$ directions, $d_{x},d_{y}$ are the pairing potential
in adjacent sites in $x$ and $y$ respectively, $\mu$ is the chemical
potential. $\psi_{i,j}(\psi_{i,j}^{\dagger})$ annihilates (creates)
a fermion at site $(i,j)$.

The Bogoliubov-de Gennes (BdG) Hamiltonian in momentum space is written
as $\mathcal{H}=\frac{1}{2}\sum_{\mathbf{k}}\Psi_{\mathbf{k}}^{\dagger}h(\mathbf{k})\Psi_{\mathbf{k}}$
up to a constant, where

\begin{gather}
h(\mathbf{k})=\left(\begin{array}{cc}
\varepsilon(\mathbf{k}) & d\left(\mathbf{k}\right)\\
d^{*}\left(\mathbf{k}\right) & -\varepsilon(\mathbf{-k})
\end{array}\right).\label{eq:H}
\end{gather}
Here, $\Psi_{\mathbf{k}}^{\dagger}=\left(\begin{array}{cc}
\psi_{\mathbf{k}}^{\dagger} & \psi_{-\mathbf{k}}\end{array}\right)$, $\mathbf{k}=(k_{x},k_{y})$, $\varepsilon(\mathbf{k})=-2t_{x}\cos(k_{x})-2t_{y}\cos(k_{y})-\mu$,
and $d\left(\mathbf{k}\right)=d_{x}\sin(k_{x})-id_{y}\sin(k_{y})$.
The distinct topological phases realized in this model as a function
of the parameters $t_{x}$, $t_{y}$, $d_{x}$, $d_{y}$, and $\mu$
have been explored in Ref. \onlinecite{asahi2012topological}. For
clarity and for later use, we review this derivation here.
\begin{figure}
\includegraphics[bb=0bp 0bp 207bp 209bp,scale=0.5]{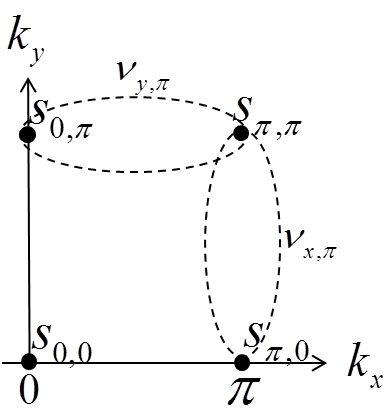}

\caption{Definition of the weak indices in the Brillouin zone {[}see Eq. (\ref{eq:nu}){]}.
The signs $s_{\mathbf{\boldsymbol{\Gamma}}_{i}}=\pm1$ are determine
by the Pfaffians of the Hamiltonian at the high symmetry points. The
weak indices $\nu_{x/y,0/\pi}$ are determined by the products of
pairs of $s_{\mathbf{\boldsymbol{\Gamma}}_{i}}$'s. Any topological
phase can be characterized by a pair of weak indices (one in each
direction), and a strong index, $\nu$.. The dashed ellipses highlight
the two weak indices chosen to label the phases in this work.\label{fig:Weak-induces-in}}
\end{figure}

The spectrum of $h(\mathbf{k})$ in (\ref{eq:H}) is $E\left(\mathbf{k}\right)=\pm\sqrt{\varepsilon^{2}(\mathbf{k})+\left|d\left(\mathbf{k}\right)\right|^{2}}$.
Assuming that $E\left(\mathbf{k}\right)\ne0$ for all $\mathbf{k}$,
i.e. the system is fully gapped, we can determine the topological
phase of the system by examining the high symmetry points that satisfy
$-\mathbf{\boldsymbol{\Gamma}}_{i}=\mathbf{\boldsymbol{\Gamma}}_{i}+\mathbf{G}$,
where $\mathbf{G}$ is a reciprocal lattice vectors. The properties
of these points $\mathbf{\boldsymbol{\Gamma}}_{i}$ will help in determining
the weak and strong topological indices that characterize the system.

The BdG Hamiltonian satisfies $Ch\left(\mathbf{k}\right)C^{-1}=-h\left(-\mathbf{k}\right)$,
where $C$ is the particle-hole transformation operator defined as $C=\tau_{x}\mathcal{K}$
($\mathcal{K}$ is complex conjugation and $\tau_{x}$ is Pauli
matrix in particle-hole space). At the high symmetry points, this
reduces to $\tau_{x}h\left(\mathbf{\boldsymbol{\Gamma}}_{i}\right)\tau_{x}=-h^{T}\left(\mathbf{\boldsymbol{\Gamma}}_{i}\right)$.
Using this relation, we can show that the transformed Hamiltonian
$\tilde{h}\left(\mathbf{\boldsymbol{\Gamma}}_{i}\right)=Uh\left(\mathbf{\boldsymbol{\Gamma}}_{i}\right)U^{-1}$
where $U=e^{i\pi\tau_{x}/4}$ is antisymmetric, $\tilde{h}\left(\mathbf{\boldsymbol{\Gamma}}_{i}\right)=-\tilde{h}\left(-\mathbf{\boldsymbol{\Gamma}}_{i}\right)^{T}=-\tilde{h}\left(\mathbf{\boldsymbol{\Gamma}}_{i}+\boldsymbol{G}\right)^{T}=-\tilde{h}\left(\mathbf{\boldsymbol{\Gamma}}_{i}\right)^{T}$.
One can therefore define 
$s_{\boldsymbol{\Gamma}_{i}}=\mbox{sign}\left\{ i\mbox{Pf}\left[\tilde{h}\left(\mathbf{\Gamma_{i}}\right)\right]\right\} $.
It is convenient to define four topological indices

\begin{align}
(-1)^{\nu_{y,\pi}} & =s_{(0,\pi)}s_{(\pi,\pi)}\nonumber \\
(-1)^{\nu_{y,0}} & =s_{(0,0)}s_{(\pi,0)}\nonumber \\
(-1)^{\nu_{x,\pi}} & =s_{(\pi,0)}s_{(\pi,\pi)}\nonumber \\
(-1)^{\nu_{x,0}} & =s_{(0,0)}s_{(0,\pi),}\label{eq:nu}
\end{align}
where $\nu_{\alpha,K}$ ($\alpha=x,y$, $K=0,\pi$) are topological invariants
of an effective 1D system in class $D$ \cite{kitaev2001unpaired}
with fixed $k_{\alpha}=K$. Fig. \ref{fig:Weak-induces-in} illustrates
the high symmetry points in the Brillouin Zone and the relations between
the topological weak indices.

\begin{figure}[H]
\begin{raggedright}
\includegraphics[bb=0bp 0bp 659.60000000000002bp 604.70000000000005bp,scale=0.3]{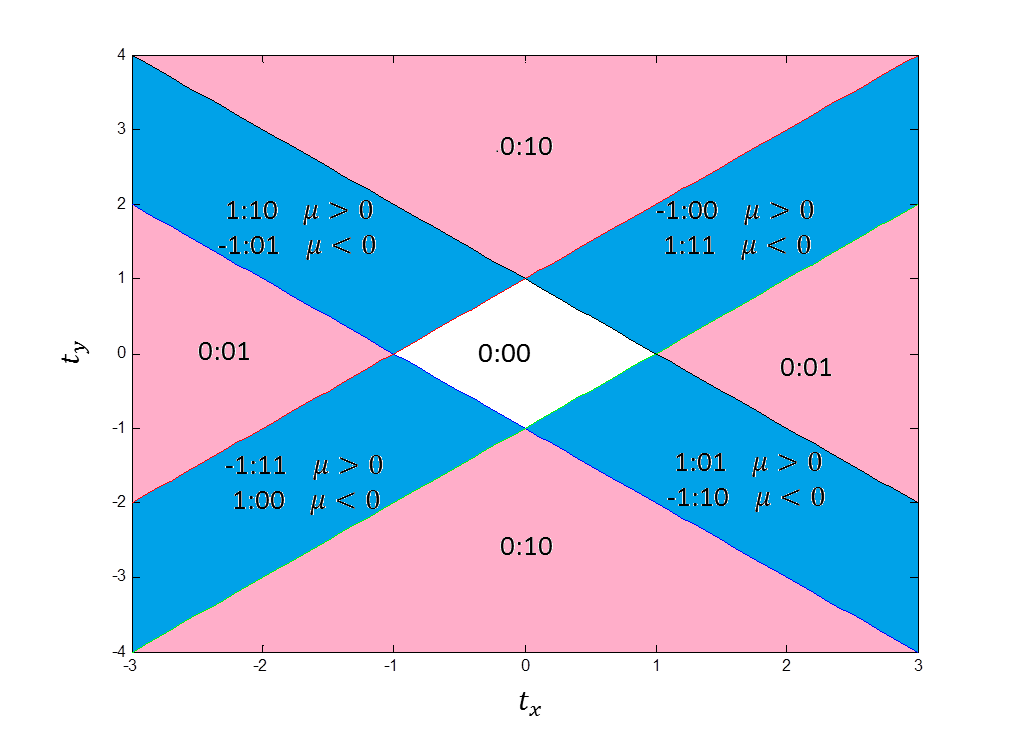}
\par\end{raggedright}

\caption{Phase diagram of the model defined in Eq.(\ref{eq: spinlees Hij})
in the $t_{x}-t_{y}$ plane. The chemical potential was set to $\mu=2$.
The topological phases are characterized by a strong index and two
weak indices $\nu\mbox{:}\nu_{x,\pi}\nu_{y,\pi}$ {[}see Eqs.(\ref{eq:nu})
and (\ref{eq:Chern}){]}. Along the phase boundaries the energy gap
closes. Strong topological phases (where $\nu\ne0$) are indicated
in blue, weak phases in pink, and the trivial phase appears in white.
  \label{fig:pd-spinless} }
\end{figure}

In addition to the weak indices $\nu_{\alpha,K}$, one can introduce
the ``strong index'' (or Chern number) $\nu$ given by\cite{TKNN}

\begin{equation}
\nu=\frac{1}{\pi}\underset{n}{\sum}\int\int dk_{x}dk_{y}\mathrm{Im}\langle\partial_{k_{x}}\psi_{n}\vert\partial_{k_{y}}\psi_{n}\rangle,\label{eq:Chern}
\end{equation}
where $\psi_{n}$ are the eigenstates of the Hamiltonian (\ref{eq:H}),
and the sum runs over the negative energy bands. \textcolor{black}{Practically,
it can be calculated numerically, see Eq.(\ref{eq:CN using P}) in
Appendix \ref{Appendix D}. }The strong index $\nu$ is related to
the weak indices by\cite{asahi2012topological}

\begin{equation}
\nu_{x,0}+\nu{}_{x,\pi}=\nu_{y,0}+\nu{}_{y,\pi}=\nu\mbox{ mod}(2).\label{eq:weak and strong relation}
\end{equation}
Therefore, the topological properties of the system are determined
by a pair of weak indices, one with an $x$ label and another with
a $y$ label, plus the strong index. In this paper, we choose to label
the different phases by the three indices $\nu\mbox{\ensuremath{:}}\nu_{x,\pi}\nu_{y,\pi}$,
where $\nu\in\mathbb{Z}$ and $\nu_{x,\pi},\nu_{y,\pi}\in\mathbb{Z}_{2}$.
For the model (\ref{eq:H}), the $\mathbb{Z}_{2}$ invariants are
easy to compute, since $s_{\mathbf{\boldsymbol{\Gamma}}_{i}}=\mbox{sign}\left[\varepsilon(\mathbf{\boldsymbol{\Gamma}}_{i})\right]$.
The phase diagram as a function of $t_{x}$, $t_{y}$ appears in Fig.
\ref{fig:pd-spinless}.

\begin{figure*}
(a)\includegraphics[bb=0bp 0bp 261bp 220bp,scale=0.4]{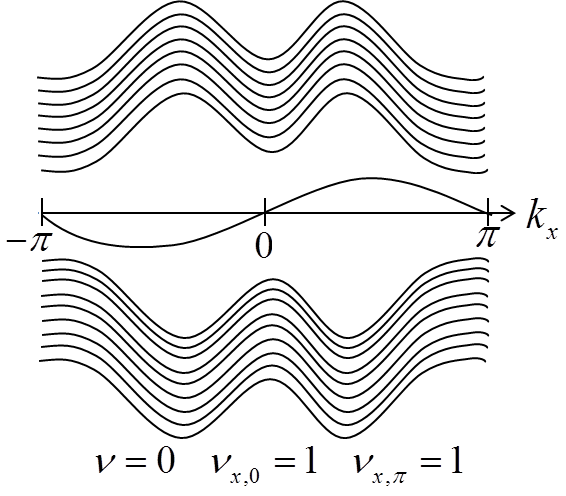}(b)
\includegraphics[bb=0bp 0bp 261bp 220.5bp,scale=0.4]{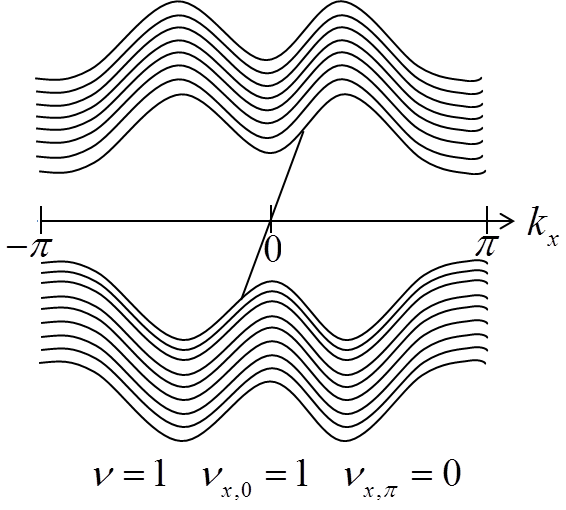}(c)\includegraphics[bb=0bp 0bp 261bp 220.5bp,scale=0.4]{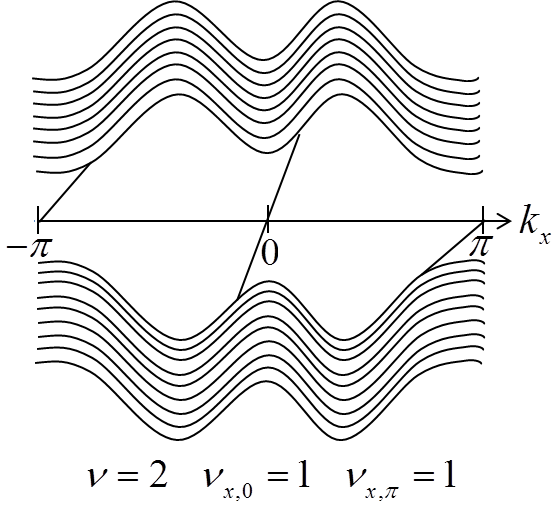}

\caption{A schematic illustration of the energy spectra of a system with an
edge parallel to $x$ as a function of $k_{x}$ in different topological
phases. (a) A phase with $\nu=0$, $\nu_{x,0}=1$, and $\nu_{x,\pi}=1$.
A band of non-chiral edge state appears, and crosses zero energy at
$k_{x}=0,\pi$. Notice that for this phase the slopes at $k_{x}=0$
and $k_{x}=\pi$ are opposite (b) A phase with $\nu=1$, $\nu_{x,0}=1$,
$\nu_{x,\pi}=0$. A chiral edge state exists, and crosses zero energy
at $k_{x}=0$. (c) A phase with $\nu=2$, $\nu_{x,0}=1$, $\nu_{x,\pi}=1$.
Two chiral edge modes with positive slopes exist. One crosses zero
energy at $k_{x}=0$ and the other at $k_{x}=\pi$.\label{fig:spectrum in v-0 and v=00003D2}}
\end{figure*}

It is well-known that the Chern number is equal to the number of
edge modes at the edge of the system, weighted by their chirality.
These chiral edge modes are robust to any weak perturbations, and
do not rely on any particular symmetry. In this sense, a phase with
a non-zero Chern number is a \emph{strong topological superconducting
phase}\textit{\emph{,}} and the Chern number is a strong topological
index. The other indices that characterize the 2D system, $\nu_{x,\pi}$
and $\nu_{y,\pi}$, are only well defined in the presence of translational
invariance in the $x$ and $y$ directions, respectively, and will
be referred to as \emph{weak topological indices. }\textit{\emph{If
at least one of the weak indices is non-zero while the Chern number
is zero, the system}}\emph{ }\textit{\emph{is in}}\emph{ }\textit{\emph{a}}\emph{
weak topological phase.}

The weak indices can be used to predict certain features of the energy
spectrum of gapless edge states that appear on boundaries in specific
directions. For example, a straight boundary parallel to the $x$
axis that preserves translational invariance in the $x$ direction
with non-zero $\nu_{x,0}$ ($\nu_{x,\pi}$) has a zero energy Majorana edge states at $k_{x}=0$ ($k_{x}=\pi$),
respectively. Using these properties, it is easy to understand Eq.(\ref{eq:weak and strong relation})
which relates the weak and strong indices. For example, the two systems
whose spectra appear in Fig. \ref{fig:spectrum in v-0 and v=00003D2}a
and \ref{fig:spectrum in v-0 and v=00003D2}c have the same Chern
number parity, because there are two edge modes, one at $k_{x}=0$
and the other at $k_{x}=\pi$. Yet, in \ref{fig:spectrum in v-0 and v=00003D2}a
the edge states have opposite chirality (opposite sign to the slope
of the edge state), hence the Chern number is zero. On the other hand,
in \ref{fig:spectrum in v-0 and v=00003D2}c the edge states have
the same chirality. Therefore, the Chern number is $2$. Fig. \ref{fig:spectrum in v-0 and v=00003D2}b
shows a case where there is only one edge state at $k_{x}=0$, hence
the parity of the number of edge states is $1$ and the Chern number
is $1$.

\section{{\normalsize{Array of quantum wires coupled to a superconducting
substrate}}}

\label{sec:spinfull model}In this section, we will discuss a more
realistic model that gives rise to the phases described in Sec. \ref{sec: Overview  Topological  supercondu}.
The first subsection (\ref{sub:Setup-and-model Sec.III}) will be
devoted to a description of the setup, and the second subsection (\ref{sub:Phase-diagram})
to the analysis of the distinct phases arise in the model as a function
of the model parameters.

\subsection{\label{sub:Setup-and-model Sec.III}Setup and model}

We envision an array of $N$ parallel quantum wires with strong spin-orbit
coupling, proximity coupled to a superconductor (Fig. \ref{fig:A-schematic-picture of the SF-1}).
Each wire has a single (spin-unresolved) mode, a large g-factor and
strong Rashba spin orbit coupling (this can be achieved, e.g., in
InAs or InSb wires\cite{das2012zero,mourik2012signatures,lutchyn2011search}).
The superconducting substrate induces a proximity gap in the wires.
It also allows electrons to tunnel relatively easily from one wire
to the next. The system is described by the following Hamiltonian:

\begin{equation}
\mathcal{H}=\mathcal{H^{\parallel}}+\mathcal{H^{\perp}}.\label{eq:Hwires}
\end{equation}
Here, $\mathcal{H}^{\parallel}$, and $\mathcal{H}^{\perp}$ describe
the intra-wire and inter-wire Hamiltonian respectively.

The intra-wire Hamiltonian, $\mathcal{H}^{\parallel}$ is given by

\begin{multline}
\mathcal{H^{\parallel}}=\sum_{j=1}^{N}\mathcal{H}_{j},\\
\mathcal{H}_{j}=\int dk_{x}[\varepsilon_{j}(k_{x})\psi_{k_{x},j}^{\dagger}\psi_{k_{x},j}+\alpha\sin(k_{x})\psi_{k_{x},j}^{\dagger}\sigma_{y}\psi_{k_{x},j}\\
-V_{z}\psi_{k_{x},j}^{\dagger}\sigma_{z}\psi_{k_{x},j}+\Delta\psi_{k_{x},j}^{\dagger}(i\sigma_{y})\psi_{-k_{x},j}^{\dagger}+h.c.].\label{eq:Hparalel}
\end{multline}
$\mathcal{H}_{j}$ is the Hamiltonian of the $j$th wire,\textcolor{black}{{}
where $\varepsilon_{j}(k_{x})=-2t_{x}\cos(k_{x})-\mu$ is t}he dispersion
along the wire ($x$ is chosen to be along the wires, see Fig. \ref{fig:A-schematic-picture of the SF-1}),
$t_{x}$ is the hopping matrix element along the wire, $\mu$ is the
chemical potential, $\alpha$ is a Rashba spin-orbit coupling term
originating from an electric field perpendicular to the wires (which
we define as the $\hat{z}$ direction), and $V_{z}$ is a Zeeman field
along $\hat{z}$. The Pauli matrices $\vec{\sigma}$ acts in spin
space. We have assumed that there is a periodic lattice along the
wires, and $\Delta$ is the pairing potential induced by the s-wave
superconductor. For now, we ignore the orbital effect of the magnetic
field; we will consider it in Sec. \ref{sec:Experimental-signature}.

The inter-wire Hamiltonian $\mathcal{H^{\perp}}$ is given by:

\begin{multline}
\mathcal{H^{\perp}}=\sum_{j=1}^{N-1}\int dk_{x}[-t_{y}\psi_{k_{x},j}^{\dagger}\psi_{k_{x},j+1}\\
-i\beta\psi_{k_{x},j}^{\dagger}\sigma_{x}\psi_{k_{x},j+1}\\
+\Delta_{y}\psi_{k_{x},j}^{\dagger}(i\sigma_{y})\psi_{-k_{x},j+1}^{\dagger}+h.c.],\label{eq:Hperp}
\end{multline}
where $t_{y}$ is the inter-wire hopping matrix element, $\Delta_{y}$
is the pairing potential associated with a process where a Cooper
pair in the superconductor dissociate into one electron in the $j$th
wire and another in the $(j+1)$th wire, and $\beta$ is the coefficient
of a spin-orbit interaction that originates from inter-wire hopping.

\textcolor{black}{Let us briefly discuss the typical magnitudes of
the parameters in Eqs.(\ref{eq:Hparalel}) and, (\ref{eq:Hperp}).
The hopping matrix element $t_{x}$ is a quarter of the bandwidth
of the conduction band in the quantum wires, and is therefore of the
order of a few electron-volts. In experimental setups similar to those
of Refs. \onlinecite{das2012zero, deng2012anomalous, mourik2012signatures},
the parameters $\left|\Delta\right|,$ $\left|V_{z}\right|$, $\left|\alpha\right|$,
and $\left|\mu+2t_{x}\right|$ (the chemical potential measured relative
to the bottom of the conduction band) are all of the order of a 0.1-1meV.
Therefore, in such setups, $t_{x}$ is much larger than all the other
parameters in the Hamiltonian. One can also imagine suppressing $t_{x}$
and creating a super-lattice. The super-lattice can be achieved for
example by applying a periodically modulated potential along the wires,
which would allow the ratio of $t_{x}$ to the other parameters to
be of order unity.}

The inter-wire hopping occurs through the superconducting substrate.
In order to get a significant inter-wire coupling, the distance between
the wires must be at most of order $\xi$, the coherence length in
the s-wave superconductor. \textcolor{black}{The inter-wire spin orbit
coupling term $\beta$ depends, mostly on the properties of the material
creating the coupling; in the case of nearly touching wires or ribbons
this term will depend mostly on the semiconducting material of the
wire, such as InAs or InSb. However, when there is a significant distance
between the wires, $\beta$ depends mostly on properties of the superconductor;
therefore, if the superconductor is made of a light element (such
as Al), $\beta$ might be negligible. To get large values of $\beta$,
one would have to use a superconductor made of a heavy element, e.g.
Pb. As we will show below, the physics depends crucially on $\beta$;
if $\beta=0$, one can not obtain gaped chiral superconducting phases.}\textcolor{red}{}

\subsection{\label{sub:Phase-diagram}Phase diagram}

We now turn to analyze the phase diagram of the model of Eq.(\ref{eq:Hwires}).
In the limit of decoupled wires, $t_{y}=\Delta_{y}=\beta=0$, this
is precisely the model studied in Refs. \onlinecite{lutchyn2010majorana}
and \onlinecite{oreg2010helical}. The phase diagram of each wire
consists of two phases, a trivial phase which is realized when $V_{z}<\sqrt{\Delta^{2}+(\mu+2t_{x})^{2}}$,
and a topological phase for $V_{z}>\sqrt{\Delta^{2}+(\mu+2t_{x})^{2}}$.
The topological phase is characterized by a zero energy Majorana mode
at the two ends of each wire\cite{lutchyn2010majorana,oreg2010helical}.
In terms of the two-dimensional topological indices described above,
the trivial phase corresponds to $\mbox{\ensuremath{\ensuremath{\nu}=0\mbox{\mbox{:}}\ensuremath{\nu_{x,\pi}}=0,\ensuremath{\nu_{y,\pi}}=0}}$,
while the non-trivial phase is a weak topological superconducting
phase labeled as $\nu=0\mbox{:}\nu_{x,\pi}=0,\nu_{y,\pi}=1$.

Next, let us consider the effect of inter-wire coupling. We will study
the phase diagram as a function of the chemical potential $\mu$ and
the Zeeman field $V_{z}$ for a fixed value of $\Delta$. Imagine
starting deep in either the $0\mbox{:}00$ or the $0\mbox{:}01$ phase,
and turning on a small inter-wire coupling. Clearly, the inter-wire
coupling cannot induce a phase transition as long as it is small compared
to the gap. In the vicinity of the phase transition between the $0\mbox{:}00$
and $0\mbox{:}01$ phases, however, the inter-wire coupling can give
rise to new phases.
\begin{figure}[H]
\begin{raggedright}
\includegraphics[bb=0bp 0bp 400bp 298.5bp,scale=0.58]{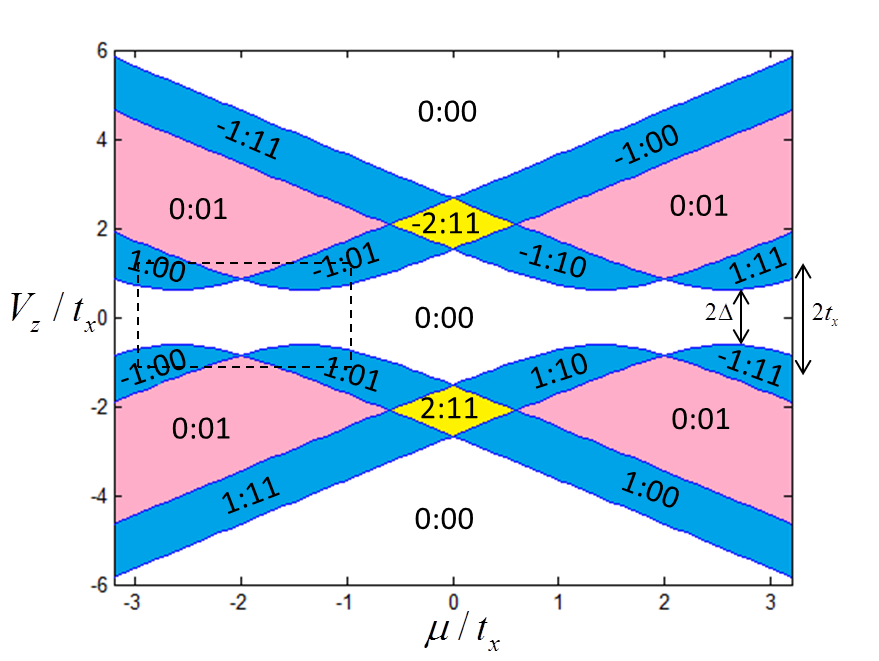}
\par\end{raggedright}

\caption{Phase diagram of a system of weakly coupled wires in proximity to
an s-wave superconductor, as a function of the Zeeman field $V_{z}$
and the chemical potential $\mu$. The topological phases are labeled
by a strong index and two weak indices, $\mbox{\ensuremath{\nu}\mbox{\mbox{:}}\ensuremath{\nu_{x,\pi}\nu_{y,\pi}}}$
{[}see Eqs.(\ref{eq:nu}), and (\ref{eq:Chern}){]}. The white regions
are topologically trivial, the pink regions are the weak topological
phases, the blue regions are the strong topological phases with a
strong index $\nu=1$, and the yellow regions are the strong topological
phases with a strong index $\nu=2$. The dashed box highlights the
region accessible in experiments using electron doped quantum wires,
in which the chemical potential is near the bottom of the conduction
band. This region \textcolor{black}{is defined by $t_{x}\gg\{\left|\Delta\right|,\left|V_{z}\right|,\left|\alpha\right|,\left|\mu+2t_{x}\right|\}$}
{[}these parameters are defined in Eqs.(\ref{eq:Hwires}), (\ref{eq:Hparalel}),
and (\ref{eq:Hperp}){]}. The parameters used in this calculation
are $t_{x}=1,$ $t_{y}=0.3$, $\Delta=0.6$, $\beta=0.3$, and $\alpha=1$.\label{fig:phase-diagram-spinful}}
\end{figure}

Fig. \ref{fig:phase-diagram-spinful} shows the phase diagram of the
model (\ref{eq:Hwires}) as a function of $\mu$ and $V_{z}$ for
fixed values of $\Delta$, $t_{x}$, $t_{y}$, $\alpha$ and $\beta$.
The phase boundaries were obtained by diagonalizing the Hamiltonian
and locating points in the $(\mu,V_{z})$ plane were the gap closes.
The spectrum of the system is given by

\begin{eqnarray}
E(\mathbf{k})^2&=&V_{z}^{2}+\Delta_{\mathrm{eff}}(\mathbf{k})^{2}+\xi(\mathbf{k}){}^{2}+|\gamma(\mathbf{k})|^{2}\nonumber \\
&\pm&2\sqrt{[V_{z}\Delta_{\mathrm{eff}}(\mathbf{k})]^{2}+(V_{z}^{2}+|\gamma(\mathbf{k})|^{2})\xi(\mathbf{k})^{2}},\label{eq:spectrum of the system}
\end{eqnarray}
where $\Delta_{\mathrm{eff}}(\mathbf{k})=\Delta+\Delta_{y}\mbox{cos}(k_{y})$
, $\xi(\mathbf{k})=-\mu-2t_{y}\mbox{cos}(k_{y})-2t_{x}\mbox{cos}(k_{x})$
and $\gamma(\mathbf{k})=\alpha i\mbox{sin}(k_{x})+\beta\mbox{sin}(k_{y})$.

The different phases are then identified by using the topological
indices of Eqs.(\ref{eq:nu}) and, (\ref{eq:Chern}). an explicit
calculation of these number is given in Appendix \ref{Appendix D}.
Slivers of phases with non-zero Chern numbers appear between the $0\mbox{:}00$
and $0\mbox{:}01$ phases. For example, examining Fig. \ref{fig:phase-diagram-spinful}
we note that upon increasing $V_{z}$ from zero at a fixed negative
value of $\mu$ between $-2$ to $-3$ (measured in units of $t_{x}$),
\textcolor{black}{the gap first closes at $\mathbf{k}=(0,0)$ and
then reopens, and a $1\mbox{:}00$ phase is stabilized. This phase
is an anisotropic realization of a chiral $p+ip$ superconductor,
and has a chiral Majorana edge mode at its boundary. Upon increasing
$V_{z}$ further, the gap at $\mbox{\textbf{k}}=(0,\pi)$ closes and
reopens, and the system enters the $0\mbox{\mbox{\mbox{\mbox{:}}}}01$
phase.}\textcolor{magenta}{{} } The points in momentum space
where the gap closes can be identified by computing the values of $s_{\mathbf{\boldsymbol{\Gamma}}_{i}}$ 
in the Brillouin Zone (See table \ref{tab:A-summery-of the phases}) and locating the point
where the sign of $s_{\mathbf{\boldsymbol{\Gamma}}_{i}}$ changes between the two neighboring phases.
\begin{figure}[H]
\begin{raggedright}
(a)
\par\end{raggedright}

\begin{raggedright}
\includegraphics[bb=80bp 200bp 550bp 570bp,scale=0.5]{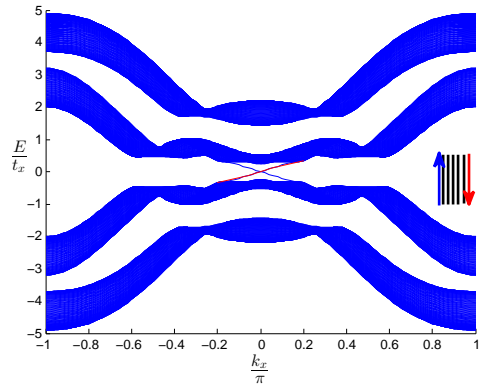}
\par\end{raggedright}

(b)

\includegraphics[bb=80bp 200bp 550bp 570bp,scale=0.5]{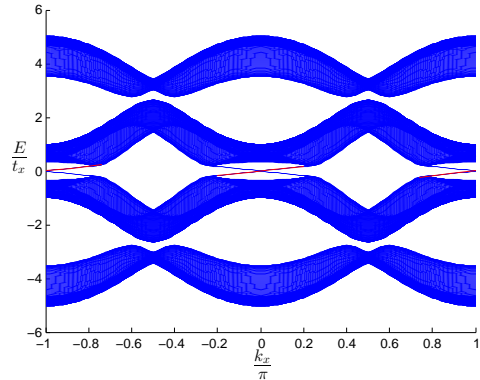}

\caption{Energy spectra of an array of coupled wires in proximity to an s-wave
superconductor, as function of the momentum along the wires, $k_{x}$.
The parameters used in the calculation are $t_{x}=1,$ $t_{y}=0.3$,
$\Delta=0.6$, $\beta=0.3$, $\alpha=1$. The system is composed of
$101$ wires, with open boundary conditions in the $y$ direction.
By varying $\mu$ and $V_{z}$, we can tune the system into different
phases: (a) For $\mu=-1.4$ and $V_{z}=0.6\sqrt{2}$, the $1\mbox{:}00$
phase is realized. This phase has one chiral mode at each edge, located
at $k_{x}=0$. The inset illustrates the edge modes in real space,
using the frame of coordinates defined in Fig. \ref{fig:A-schematic-picture of the SF-1}.
The $+x$ ($-x$) moving edge mode is colored in red (blue), respectively.
(b) For $\mu=0$ and $V_{z}=2.088$, the phase $2\mbox{:}11$ emerges
(see Sec. \ref{sub:-phase v=00003D2 }). Two chiral edge modes appear
at each edge, one at $k_{x}=0$ and the other at $k_{x}=\pi$.\label{fig:The-energy-spectrum}}
\end{figure}

\textcolor{black}{As discussed earlier, the experimentally accessible
regime in setups similar to those of Refs. \onlinecite{mourik2012signatures, deng2012anomalous, das2012zero}
is defined by $t_{x}\gg\{\left|\Delta\right|,\left|V_{z}\right|,\left|\alpha\right|,\left|\mu+2t_{x}\right|\}$.
We highlight the accessible region by a dashed box in Fig. \ref{fig:phase-diagram-spinful}.
In order to access all the possible phases in Fig. \ref{fig:phase-diagram-spinful},
one needs to suppress $t_{x}$, for example by applying a periodically
modulated potential along the wires, creating a super-lattice.}

The spectrum of a system with a finite number of wires in the $1\mbox{\mbox{:}}00$
phase is presented in Fig. \ref{fig:The-energy-spectrum}a, as a function
of momentum along the wires. As expected, there are two counter-propagating
edge modes within the bulk gap. These modes are localized on the opposite
sides of the system.

\subsubsection{\label{sub:-phase v=00003D2 }A phase with a strong index $\nu=2$}

It is interesting to note that the phase diagram (Fig. \ref{fig:phase-diagram-spinful})
contains a $2\mbox{:}11$ phase, with a Chern number $\nu=2$ and
two co-propagating chiral edge modes. This phase appears around $\mu=0$
for large Zeeman fields ($V_{z}\approx2t_{x}$). One can understand
qualitatively the emergence of this phase as follows. Focusing in
Fig. \ref{fig:phase-diagram-spinful} on the region in which $-2t_{x}<\mu<-2t_{y}$,
as the Zeeman field is increased from $V_{z}=0$, the gap closes at
$\mathbf{k}=(0,\pi)$ and reopens, stabilizing a $-1\mbox{:}01$ phase.
This phase is characterized by a chiral edge mode, which appears around
$k_{x}=0$ in a system with a boundary parallel to the $x$ axis.
Similarly, for the particle-hole conjugated path at $2t_{y}<\mu<2t_{x}$,
the gap closes and reopens at $\mathbf{k}=(\pi,0)$ upon increasing
$V_{z}$ from zero, and one finds a $-1\mbox{:}10$ phase with a chiral
edge mode around $k_{x}=\pi$ at a boundary along the $x$ axis. Near
$\mu=0$, these two gap closings coincide, and we find a phase that
has \emph{both} a chiral edge modes at $k_{x}=0$ and at $k_{x}=\pi$
(see Fig. \ref{fig:The-energy-spectrum}b). We analyze the appearance
of this $\nu=2$ phase in detail in Appendix \textbf{\ref{Appendix C}}.

\subsubsection{\label{sub:``Sweet-point''}``Sweet point'' with perfectly localized
edge states}

Interestingly, upon tuning the Zeeman field $V_{z}$, there is a special
``sweet point'' at the center of the $1\mbox{\mbox{:}}00$ phase (as
well as in the other chiral phases) in which the chiral states
at an edge parallel to the $x$ axis are almost entirely localized
on the outmost wires. \textcolor{black}{(Notice that the localization
lengths of edge states at edges along the $x$ and $y$ axes are generically
different from each other, due to the anisotropy of our system.)}
At this point, the edge states on the two opposite edges do not mix
even in systems with a small number of wires, making it attractive
from the point of view of experimental realizability. This point in
parameters space is analogous to the special point in the Kitaev\textquoteright{}s
one-dimensional chain model\cite{kitaev2001unpaired}, in which the
Majorana end states are localized on the last site. In our two-dimensional
setup, we will show how one can access this point by tuning the magnetic
field.

We now derive a criterion for realizing the ``sweet point'', and give
a simple picture for its emergence. First, let us consider a system
without coupling between the wires. The Hamiltonian of the $j$th
wire Eq.(\ref{eq:Hparalel})\textbf{ }can be written as $\mathcal{H}_{j}=\frac{1}{2}\underset{k_{x}}{\sum}\Psi_{k_{x}\mathbf{,}j}^{\dagger}h_{j}(k_{x})\Psi_{k_{x}\mathbf{,}j}$,
where

\begin{equation}
h_{j}(k_{x})=\varepsilon_{j}(k_{x})\tau_{z}-V_{z}\sigma_{z}+\alpha\sin(k_{x})\tau_{z}\sigma_{y}+\Delta\tau_{x}.\label{eq:Hj}
\end{equation}
Here, $\Psi_{k_{x},j}^{\dagger}=\left(\psi_{\uparrow,k_{x},j}^{\dagger},\psi_{\downarrow,k_{x},j}^{\dagger}\psi_{\downarrow,-k_{x},j},-\psi_{\uparrow,-k_{x},j}\right)$,
and $\vec{\tau}$ are Pauli matrices acting in Nambu (particle-hole)
space.

\begin{figure}
\begin{centering}
\includegraphics[bb=0bp 0bp 370bp 200bp,scale=0.7]{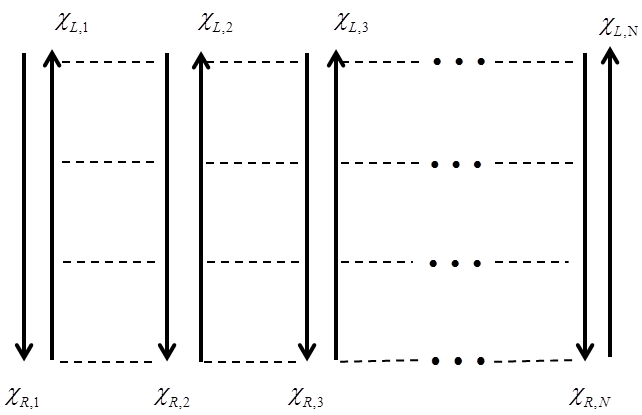}
\par\end{centering}

\caption{Schematic illustration of the physics leading to the ``sweet point''
(Sec. \ref{sub:``Sweet-point''}). At low energies, each wire has
two counter-propagating Majorana modes. When the condition in Eq.(\ref{eq:"sweat point"-1})
is satisfied, the inter-wire coupling takes a special form such that
the right moving Majorana mode of each wire couples only to the left
moving Majorana mode of the adjacent wire. As a result, this pair
of Majorana mode is gapped out, leaving only the two outmost counter-propagating
modes gapless. \label{fig:Sweat point}}
\end{figure}

The strategy in constructing the ``sweet point'' is as follows. We
first tune the parameters of the single wire Hamiltonian Eq.(\ref{eq:Hj})
to the critical point at the transition from the trivial to the topological
phase. At this point, the low-energy theory is described by two counter-propagating
Majorana modes. Turning on the inter-wire coupling induces backscattering
between these modes. At the sweet point, the inter-wire coupling takes
a special form such that the right moving Majorana mode of one wire
couples only to the left moving Majorana mode of the adjacent wire
(see Fig. \ref{fig:Sweat point}). This coupling gaps this pair of
modes out, leaving only the two outmost counter-propagating modes
gapless. This is similar to the approach of Refs. \onlinecite{kane2002fractional, teo2011luttinger, sondhi2001sliding, mong2013universal}
for constructing quantum Hall phases starting from weakly coupled
wires.

Let us demonstrate this by focusing on the single-wire critical point
at
\begin{equation}
\mu=-2t_{x}+\sqrt{V_{z}^{2}-\Delta^{2}},\label{eq:critical}
\end{equation}
in which the gap closes at $k_{x}=0$. We diagonalize the Hamiltonian
by a Bugoluibov transformation of the form
\begin{equation}
\Psi_{k_{x},j}=W_{k_{x}}\tilde{\Psi}_{k_{x},j},\label{eq:Psi}
\end{equation}
where $\tilde{\Psi}_{k_{x},j}^{\dagger}=\left(\psi_{2,k_{x},j}^{\dagger},\psi_{1,k_{x},j}^{\dagger},\psi_{1,k_{x},j},-\psi_{2,k_{x},j}\right)$,
and the matrix $W_{k_{x}}$ is given by:

\begin{equation}
W_{k_{x}=0}^{\dagger}=\sqrt{\frac{\left|\Delta\right|}{2V_{z}s}}\left(\begin{array}{cccc}
s & 0 & 1 & 0\\
-1 & 0 & s & 0\\
0 & s & 0 & 1\\
0 & -1 & 0 & s
\end{array}\right).\label{eq:W}
\end{equation}
Here, $s\equiv\frac{-2t_{x}-\mu+V_{z}}{\Delta}$. At the critical
point, $\psi_{1,k_{x},j}$ is gapless and disperses linearly, while
$\psi_{2,k_{x},j}$ remains gapped. Expanding near $k_{x}=0$, the
Hamiltonian takes the form

\begin{equation}
\mathcal{H^{\parallel}}=\sum_{k_{x},j}\left[vk_{x}\psi_{1,k_{x},j}^{\dagger}\psi_{1,k_{x},j}+h\psi_{2,k_{x},j}^{\dagger}\psi_{2,k_{x},j}+O(k_{x}^{2})\right],
\end{equation}
where \textcolor{black}{$v=\frac{\Delta\alpha}{V_{z}}$}. Inserting
Eq.(\ref{eq:Psi}) into Eq.(\ref{eq:Hperp}), and using the explicit
form of $W_{k_{x}}$ given in Eq.(\ref{eq:W}), the inter-wire coupling
Hamiltonian projected onto the low-energy ($\psi_{1,k_{x},j}$) sector
becomes

\begin{align}
\mathcal{H}^{\perp} & =\underset{k_{x},j}{\sum}\left[\left(t_{y}(s^{2}-1)-2\Delta_{y}s\right)\psi_{1,k_{x},j}^{\dagger}\psi_{1,k_{x},j+1}\right.\nonumber \\
 & +\left.i2s\beta\psi_{1,k_{x},j}^{\dagger}\psi_{1,k_{x},j+1}^{\dagger}+h.c.\right],\label{eq:Kitaev's H spinfull}
\end{align}
At $k_{x}=0$, the Hamiltonian is identical to Kitaev's one-dimensional
chain model\cite{kitaev2001unpaired} with zero chemical potential.
This model simplifies greatly for a special choice of parameters
such that $\left|t_{y}(s^{2}-1)-2\Delta_{y}s\right|=\left|2s\beta\right|$.
In terms of the physical parameters, this condition is written as

\begin{equation}
\left|t_{y}\frac{2t_{x}+\mu}{\Delta}+\Delta_{y}\right|=\left|\beta\right|.\label{eq:"sweat point"-1}
\end{equation}
For these parameters, the $k_{x}=0$ Hamiltonian is easily diagonalized
by introducing Majorana fields

\begin{align}
\chi_{R,j} & =\psi_{1,j}e^{i\phi}+\psi_{1,j}^{\dagger}e^{-i\phi}\nonumber \\
\chi_{L,j} & =-i\psi_{1,j}e^{i\phi}+i\psi_{1,j}^{\dagger}e^{-i\phi}.
\end{align}
Here, \textcolor{black}{$\phi=\frac{1}{2}\mbox{Arg\ensuremath{\left(\frac{i2s\beta}{t_{y}(s^{2}-1)-2\Delta_{y}s}\right)}=\ensuremath{\frac{1}{2}}Arg\ensuremath{\left(\frac{-i\beta}{\frac{t_{y}(2t_{x}+\mu)}{\Delta}+\Delta_{y}}\right)}}$.}\textcolor{red}{{}
}\textcolor{black}{In terms of these fields, the Hamiltonian takes
the form}
\textcolor{black}{
\begin{equation}
\mathcal{H^{\perp}}=i\underset{j}{\sum}|2s\beta|\chi_{L,j}\chi_{R,j+1}.
\end{equation}
}
The resulting phase has two chiral edge modes on the two opposite
edges, which are completely localized on the outmost wires, up to
corrections of the order of $\frac{t_{y}}{V_{z}}$ due to virtual
excitation to the gapped mode $\psi_{2,k_{x}}$. If the condition
in Eq.(\ref{eq:"sweat point"-1}) is not exactly 	satisfied, the edge
states become more spread out in the direction perpendicular to the
wires, but remain localized near the boundary as long as the bulk
gap does not close.
\begin{figure}[H]
\raggedright{}\includegraphics[scale=0.4]
{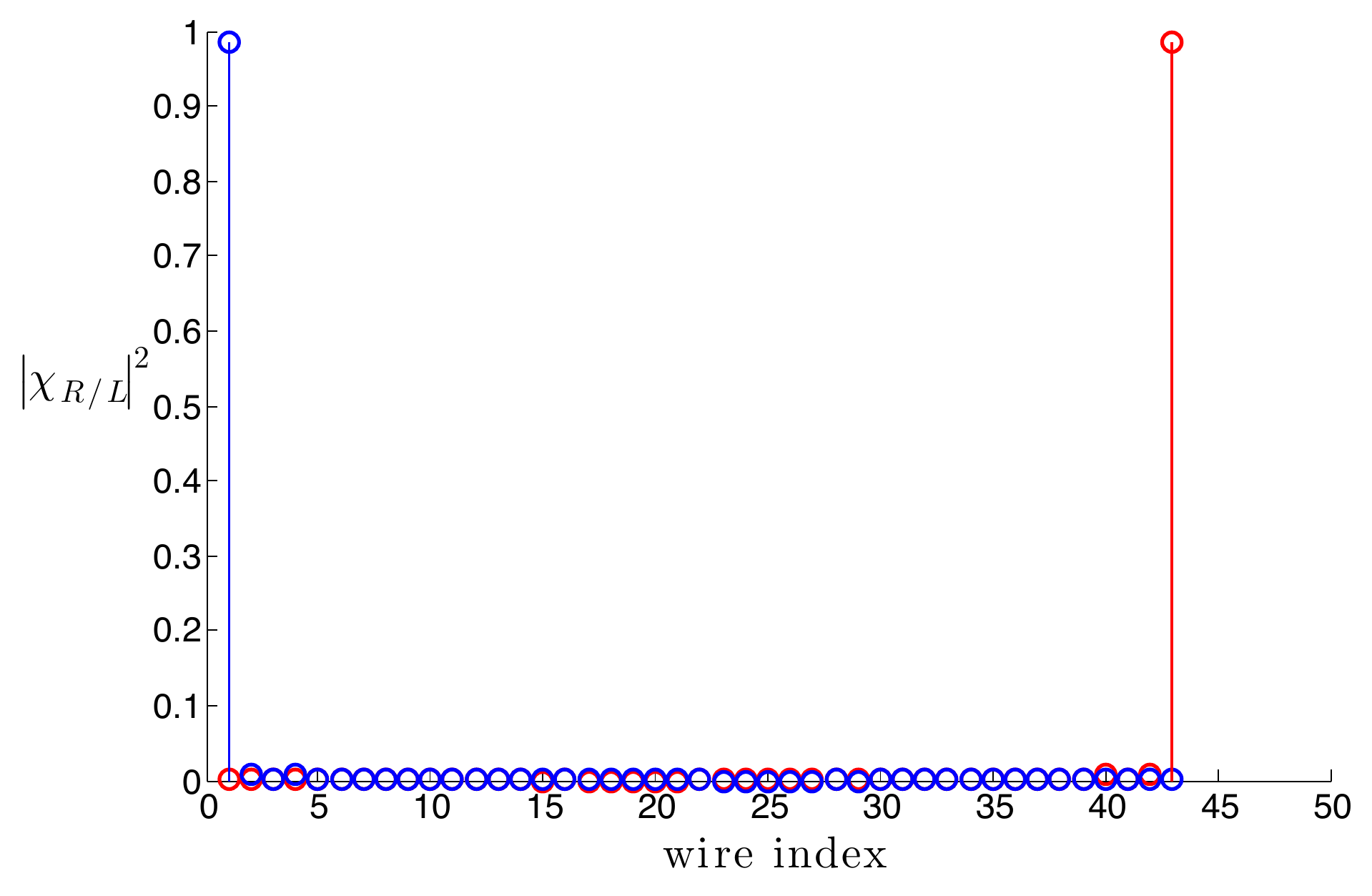}\caption{Wave functions of the two Majorana edge modes in a system of 43 wires
tuned to the ``sweet point'' {[}Eq.~(\ref{eq:"sweat point"-1}){]}.
We plot the probability distributions of the lowest energy states
at $k_{x}=0$, as a function of the wire index along $y$. The two
zero energy states are almost fully localized at the first wire (blue)
and on the last wire (red). The following parameters were used: $t_{x}=0.5$,
$t_{y}=0.1$, $\Delta=0.6$, $\Delta_{y}=0$, $\mu=-1.4$, $V_{z}=0.6\sqrt{2}$,
$\beta=0.1$, and $\alpha=1$. \label{fig:probability of the loest energy stste in the sweat point-1}}
\end{figure}

One can tune into the ``sweet point'' by setting $V_{z}$ and $\mu$
such that both Eqs.(\ref{eq:critical}) and (\ref{eq:"sweat point"-1})
are satisfied. The required parameters are $V_{z}=\pm\sqrt{\Delta^{2}+(\mu+2t_{x})^{2}}$
and $\mu=-2t_{x}-\frac{|\Delta|}{t_{y}}(|\Delta_{y}|\pm|\beta|)$.

We tested the ``sweet point'' numerically, by diagonalizing the
Hamiltonian (\ref{eq:Hwires}) for a system with a finite number of
wires. In Fig. \ref{fig:probability of the loest energy stste in the sweat point-1},
we present the wave functions of the lowest energy states as a function
of position perpendicular to the wires. As expected, the wave functions
of these states is almost localized on the outmost wires.

\subsubsection{In-plane Zeeman magnetic field }

Applying an in-plane magnetic (Zeeman) field provides an additional
experimentally accessible knob to tune the system between different
phases. We now consider its effect on the phase diagram. Note that
in our system, a perpendicular magnetic field is essential in order
to realize the strong topological phase (this is different from the
case considered in Ref. \onlinecite{alicea2010majorana}, due to the
different form of the spin-orbit coupling). The in-plane magnetic
field generally destroys the topological phases, leading to a gapless
phase instead.

\begin{figure}
\includegraphics[bb=30bp 0bp 579bp 310bp,scale=0.5]{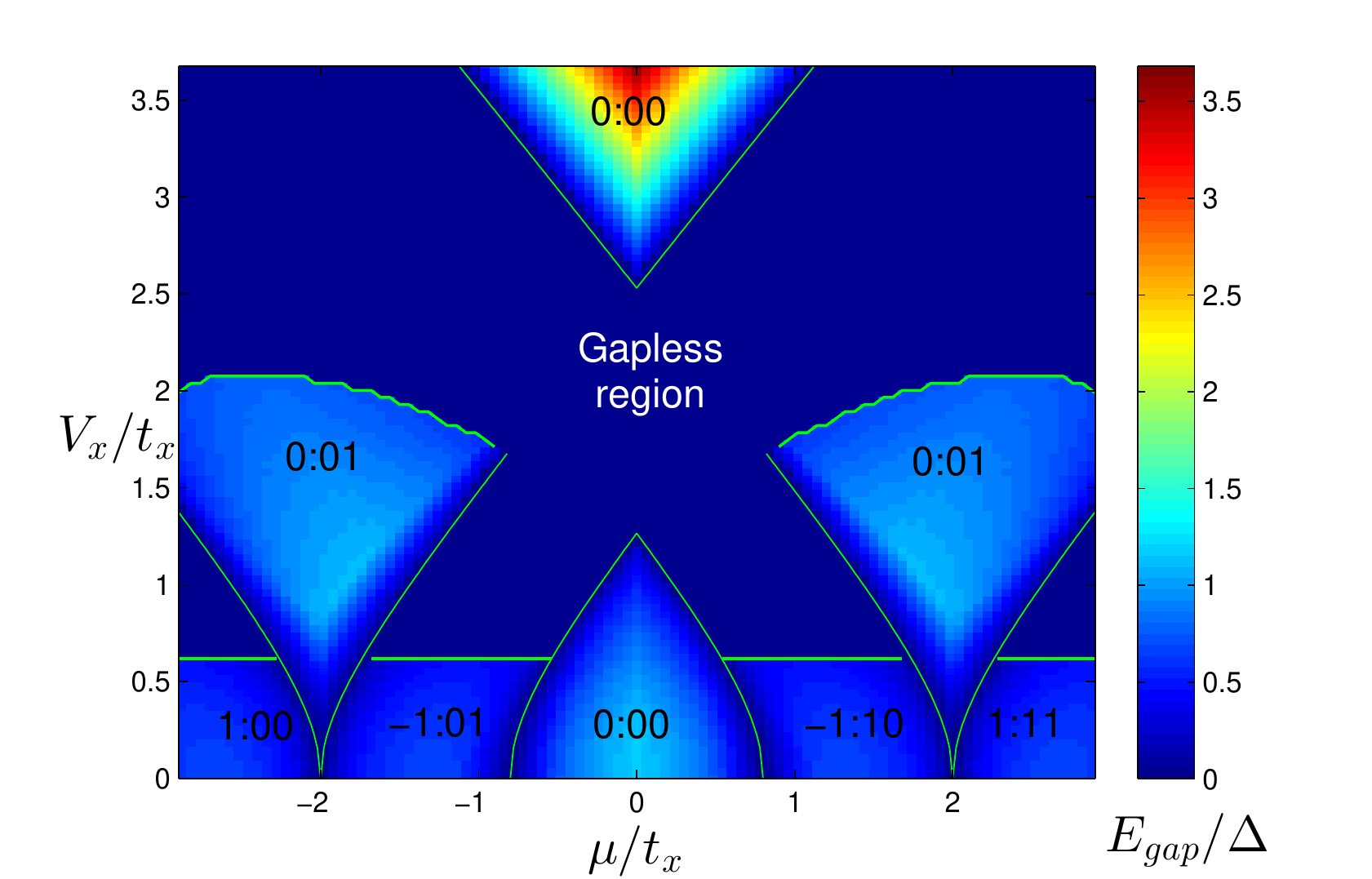}

\caption{The phase diagram and the energy gap as a function of an in plane
Zeeman field $V_{x}$ and chemical potential $\mu$, at a fixed $V_{z}$.
The light green line marks the phase boundaries, along which the gap
closes. Notice that the $V_{x}=0$ line corresponds to a constant
$V_{z}$ cut in Fig. \ref{fig:phase-diagram-spinful}. The parameters
used in this calculation are: $t_{x}=1\mbox{, }t_{y}=0.3\mbox{, }\Delta=0.6\mbox{, }\beta=0.3\mbox{, \ensuremath{V_{z}=0.6\sqrt{2}}and }\alpha=1$.\label{fig:The-minimum-energy gap Vx-mu}}
\end{figure}

In the presence of an in-plane Zeeman field applied parallel to the
wires, we should add the term $-V_{x}\psi_{k_{x},j}^{\dagger}\sigma_{x}\psi_{k_{x},j}^{\dagger}$
to Eq.(\ref{eq:Hparalel}). Then, the spectrum is given by
\begin{multline}
E^{2}(\mathbf{k})=V_{\mathrm{tot}}^{2}+\Delta_{\mathrm{eff}}^{2}+\xi(\mathbf{k}){}^{2}+|\gamma(\mathbf{k})|^{2}\\
\pm2\sqrt{V_{\mathrm{tot}}^{2}\Delta_{\mathrm{eff}}^{2}+(V_{\mathrm{tot}}^{2}+|\gamma(\mathbf{k})|^{2})\xi(\mathbf{k})^{2}+(V_{x}\mbox{sin}(k_{y})\beta)^{2}},
\end{multline}
where $V_{\mathrm{tot}}=\sqrt{V_{z}^{2}+V_{x}^{2}}$.
The condition for a closer of the gap is:
\begin{equation}
V_{z}^{2}+V_{x}^{2}=(\mu\mp2t_{x}\mp2t_{y})^{2}+\Delta^{2}.\label{eq:closer of the gap (with Vy)}
\end{equation}
Fig. \ref{fig:The-minimum-energy gap Vx-mu} shows the phase diagram
as a function of $\mu$ and $V_{x}$, fixing $V_{z}=\sqrt{2}\Delta$.
The line $V_{x}=0$ corresponds to a line of fixed $V_{z}$ of
the phase diagram shown in Fig. \ref{fig:phase-diagram-spinful}.
Upon raising $V_{x}$, a gapless (metallic) region is formed. The
gap closes because of the destruction of the proximity effect by the
in-plane field, due to the Zeeman shift of the normal state energy
at $\mathbf{k}$ relative to $-\mathbf{k}$. \textcolor{black}{The}\textcolor{red}{{}
}\textcolor{black}{effect of an in-plane field perpendicular to the
wires ($V_{y}$) is qualitatively similar, but the ``bubbles'' of
the $0\mbox{:}01$ phase do not appear, and are replaced by gapless
regions. (Notice that the response for a Zeeman field in the $x$
and $y$ directions is different because of the anisotropy of our
system.)}

One can also show that the ``sweet point'' within the strong topological
phases survives in the presence of an in-plane field. The sweet point
condition is given by Eqs.(\ref{eq:"sweat point"-1}) and (\ref{eq:closer of the gap (with Vy)})\textcolor{black}{.}

\section{The orbital effect of the magnetic field in a 2D p-wave superconductor}

\label{sec:The orbital effect}

So far, we have neglected the orbital effect of the magnetic field,
treating only the Zeeman effect. This assumption is justified in the
limit of large g-factor, $g\gg 1$. In this section, we will consider the
orbital effects of the magnetic field.

We begin by discussing the condition for the appearance of vortices in
our system. We assume that the s-wave superconductor is a narrow strip,
whose width $d$ is small compared to its length and to the bulk penetration
length. Under these conditions, the critical field for creating a
single vortex in the superconductor is\cite{stan2004critical} $H_{\mathrm{c1}}\sim\Phi_{0}/d^{2}$,
where $\Phi_{0}=h/2e$. This gives $H_{\mathrm{c1}}\sim2mT\left(1\mu m/d\right)^{2}$
.

To get a feeling for the value of this critical field in realistic setups,
let us consider a system with $N$ wires made of InAs, similar to
those of Ref.  \onlinecite{das2012zero}. We assume that the distance
between the wires is of order $\xi$ (ensuring a reasonable inter-wire
coupling). If we take $\xi\approx40nm$ (as in Pb) and $N=25$, we
get $d\approx N\xi\approx1\mu m$ and the critical magnetic field
for creating a single vortex is $B\approx2mT$. In order to be in
the topological phase the Zeeman field must satisfy $V_{z}=\frac{g\mu_{\mathrm{B}}B}{2}>\sqrt{\Delta_{\mathrm{ind}}^{2}+(\mu+2t_{x})^{2}}$
(where $\Delta_{\mathrm{ind}}$ is the induced superconducting gap
in the wire). In InAs, $g\approx20$ and $\Delta_{\mathrm{ind}}$
can be of the order of $50\mu eV$\cite{das2012zero}. This gives
that the required magnetic field to be in the topological phase is
$B>30mT$, and thus vortices are present in the strip. As we decrease
the size of the system, the critical field increases. For a system
with $N=5$ and $d\approx0.2\mu m$, for example, the critical field
is $B\approx50mT$, and one can realize a vortex-free topological
phase.

Below, we discuss features of the quasi-particle spectrum in the presence
of an orbital field that can be used as a signature of topological
phase in the system.

\subsection{Majorana zero modes in vortex cores}

The chiral phase is characterized by the presence of a Majorana zero
mode at each vortex core\cite{read2000paired}. For an applied field
slightly above $H_{c1}$, the ground state contains vortices along
the strip, with a Majorana zero mode at each core. In addition, when
the number of vortices is odd, the chiral Majorana mode on the edge
has a mid-gap state. In general, the Majorana states in the vortex
core can leak into the edge mode. However, if we choose parameters
such that the system is near the sweet point described above, such
that the effective coherence length transverse to the wires is essentially
one inter-wire spacing, the mixing between the vortex core states
and the chiral edge modes can be made negligibly small (assuming that
the wires are sufficiently long), as can be seen in Fig. \ref{fig:The-probability-distribution with a vortex}.
One can show that the sweet point condition, Eq.(\ref{eq:"sweat point"-1}),
remains unmodified when projecting to the lower energy bands 
to leading order in the orbital magnetic field (see Appendix \ref{sec:the orbital field and the =00201Csweet point=00201D }).

\subsection{Doppler shifted chiral edge states}

\begin{figure}
\begin{raggedright}
(a)
\par\end{raggedright}

\includegraphics[bb=80bp 220bp 528bp 580bp,scale=0.4]{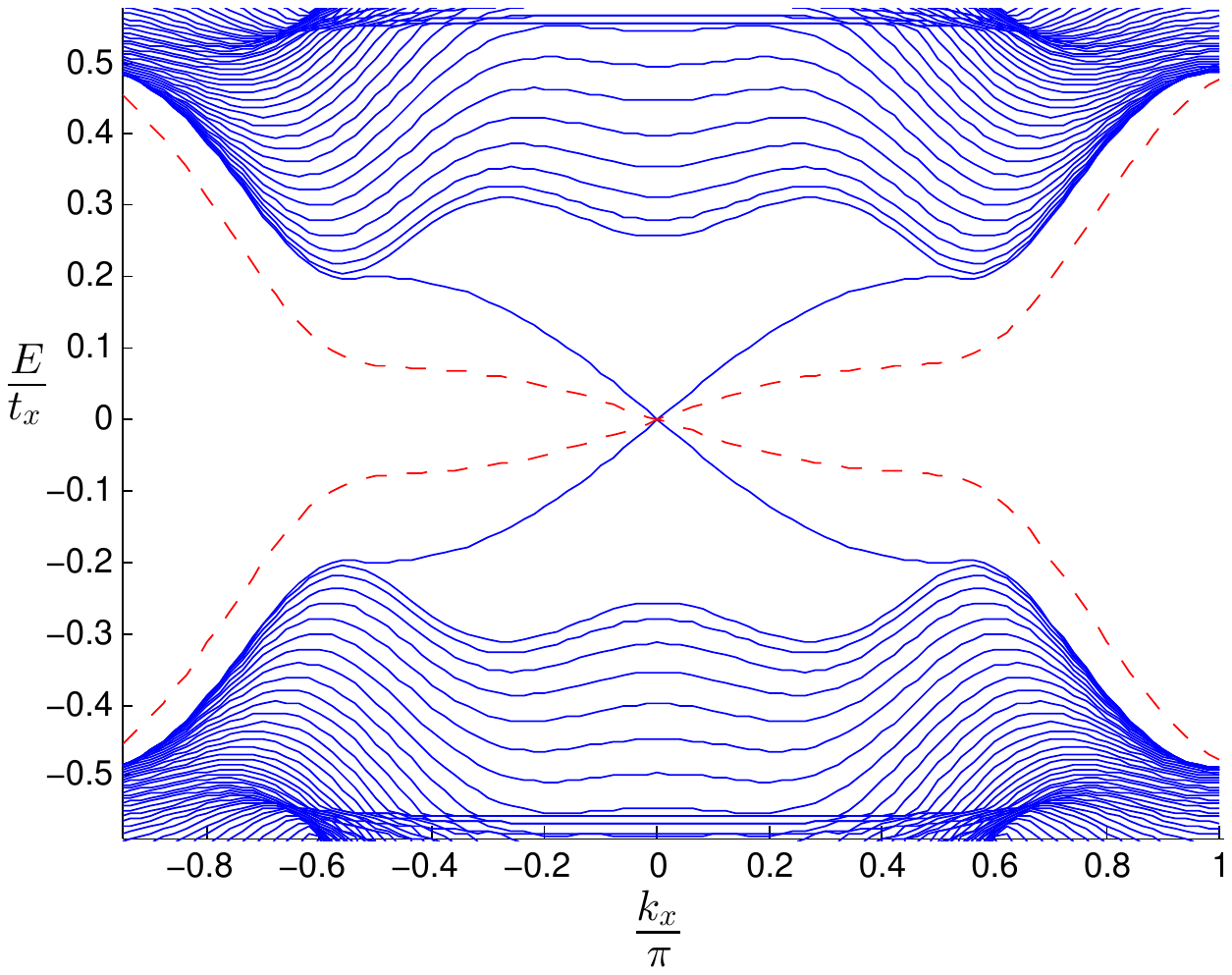}

\begin{raggedright}
(b)
\par\end{raggedright}

\includegraphics[bb=80bp 220bp 528bp 580bp,scale=0.4]{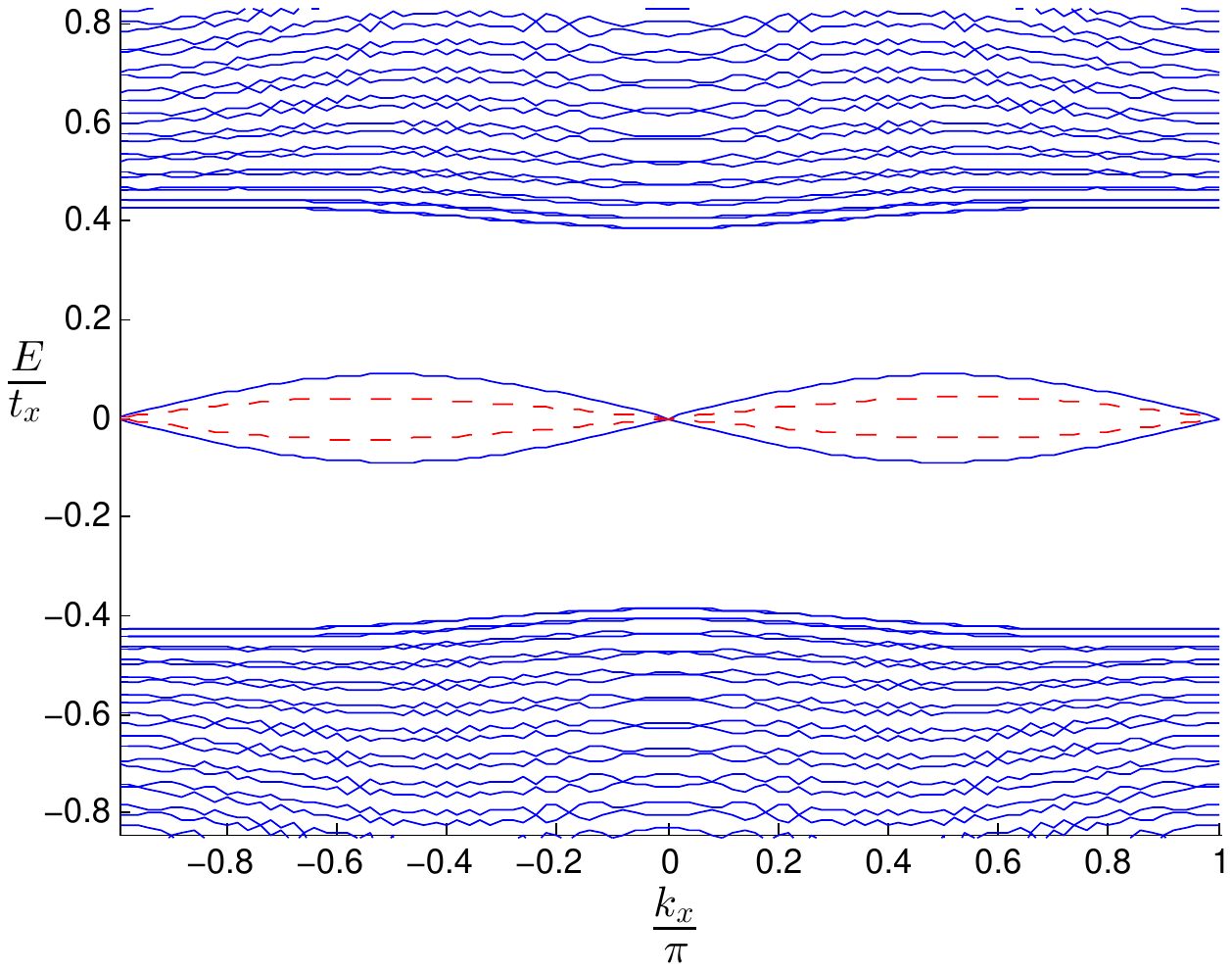}

\caption{The effect of an orbital magnetic field
on the energy spectrum in the 1:00 strong phase (a) and the 0:01 weak
phase (b). The dashed red lines are the spectra of the edge states
in the presence of an orbital field, as a function of the momentum
along the edge. The solid blue lines are the corresponding spectra
without an orbital magnetic field. The orbital field induces supercurrents
in the superconductor. As a result, the velocity of the edge modes
changes. The parameters used in the calculations are $t_{x}=1$, $t_{y}=0.3$,
$\Delta=0.6$, $\beta=0.3$, $V_{z}=0.6\sqrt{2}$, $V_{x}=0$, and
$\alpha=1$. The chemical potentials are (a) $\mu=-1.4$, and (b)
$\mu=-2$.\label{fig:Chang in the slope of the edge state}}
\end{figure}

In addition to inducing vortices, the orbital field induces circulating
orbital currents in the sample. These orbital currents modify the
low-energy density of states (DOS) due to a ``Doppler shift'' of the
quasi-particles at the edge. In a \emph{chiral} superconductor, the
Doppler shift either enhances or suppresses the DOS at the edge, depending
on whether the orbital supercurrent is parallel or anti-parallel to
the propagation direction of the chiral edge state\cite{yokoyama2008chirality}.
Using the London gauge near the edge, such that $\nabla\phi=0$ (where
$\phi$ is the phase of the order parameter), the external orbital
current is proportional to the vector potential $\vec{A}$. Consider
a system defined on the half-plane $y>0$, with an edge at $y=0$.
To linear order in $k_{x}$ and $A_{x}$, The low-energy quasi-particle
spectrum is given by

\begin{equation}
E(k_{x})=k_{x}(v_{x}+A_{x}\delta_{x}).\label{eq:energy spectrum with orbital field-1}
\end{equation}
 The velocity $v_{x}$ and the coefficient $\delta_{x}$ can be calculated
perturbatively in $k_{x}$ and $A_{x}$.
In the model described in Sec. \ref{sec:spinfull model}, the perturbative
calculation gives $v_{x}=\frac{\Delta\alpha}{V_{z}}$ and $\delta_{x}=t_{x}$.
See Appendix \ref{sec:appendix A the spectrum with an orbital field }
for an explicit derivation of this result. Since the local DOS is
proportional to the inverse of $dE/dk(k=0)$, Eq.(\ref{eq:energy spectrum with orbital field-1})
shows that the zero-energy DOS depends linearly on $A_{x}$\cite{yokoyama2008chirality}.
Fig. \ref{fig:Chang in the slope of the edge state}a shows how the
slope of the chiral edge state changes when the orbital magnetic field
is not negligible. Similarly, one can calculate the spectrum of the
edge mode at an edge parallel to $y$. For such an edge, $v_{y}=\frac{\Delta\beta}{V_{z}}$
and $\delta_{y}=t_{y}$.

Surprisingly, the linear dependence of the local DOS at the edge on
the supercurrent is not limited to the strong (chiral) topological
phases, but exists also in the weak phases. E.g., consider a system
in the $0\mbox{:}01$ phase with a straight edge parallel to $y$.
There are low-energy edge modes near $k_{y}=0$ and $k_{y}=\pi$,
whose dispersions have opposite slopes. The dispersion of the edge
mode near $k_{y}=\pi$ is given by $E(\pi+\delta k_{y})=-\delta k_{y}(v_{y}+A_{y}\delta_{y})$,
to linear order in $\delta k_{y}$. Therefore, if we apply a supercurrent
near the edge such that the slope of the edge mode at $k_{y}=0$ increases
in magnitude, the slope of the mode at $k_{y}=\pi$ increases as well,
and the total DOS at the edge decreases linearly with the current,
as can be seen in Fig. \ref{fig:Chang in the slope of the edge state}b.

In general, a linear dependence of the DOS on the supercurrent is
possible if time-reversal symmetry is broken. In our system, time-reversal
is broken by the external magnetic field, which is present both in
the weak and the strong topological phases. In the weak phases, the
edge modes do not carry current; nevertheless, a supercurrent couples
to the edge modes through the phase of the condensate.
\textcolor{red}{ }

In the presence of vortices, the orbital effect leads to an interesting
variation of the low-energy local DOS at the edge. Each vortex produces
circulating supercurrent. Therefore, the superfluid velocity at the
edge varies as a function of position; it is either enhanced or suppressed
in regions of the edge which are close to a vortex core, depending
on the chirality of the vortex relative to that of the superconductor.
(In our system, the relative chirality of the vortices and the superconductor
depends on the signs of the spin-orbit coupling terms $\alpha$ and
$\beta$, and is not easy to control externally.) Therefore, according
to Eq.(\ref{eq:energy spectrum with orbital field-1}), the local
DOS at the edge shows either a dip or a peak in the vicinity of a
vortex in the bulk.
\begin{figure*}
(a)\includegraphics[scale=0.5]{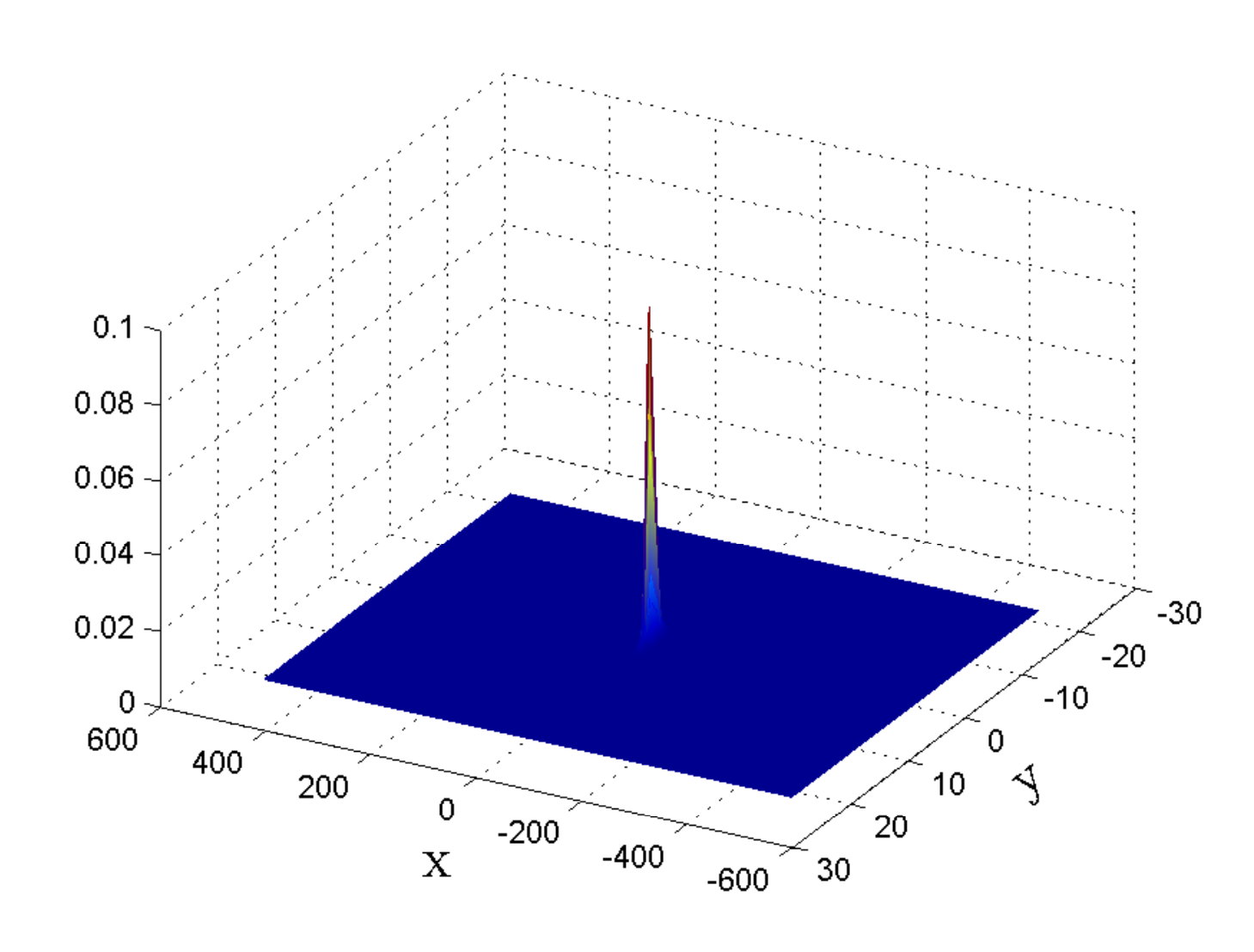}(b)\includegraphics[scale=0.5]{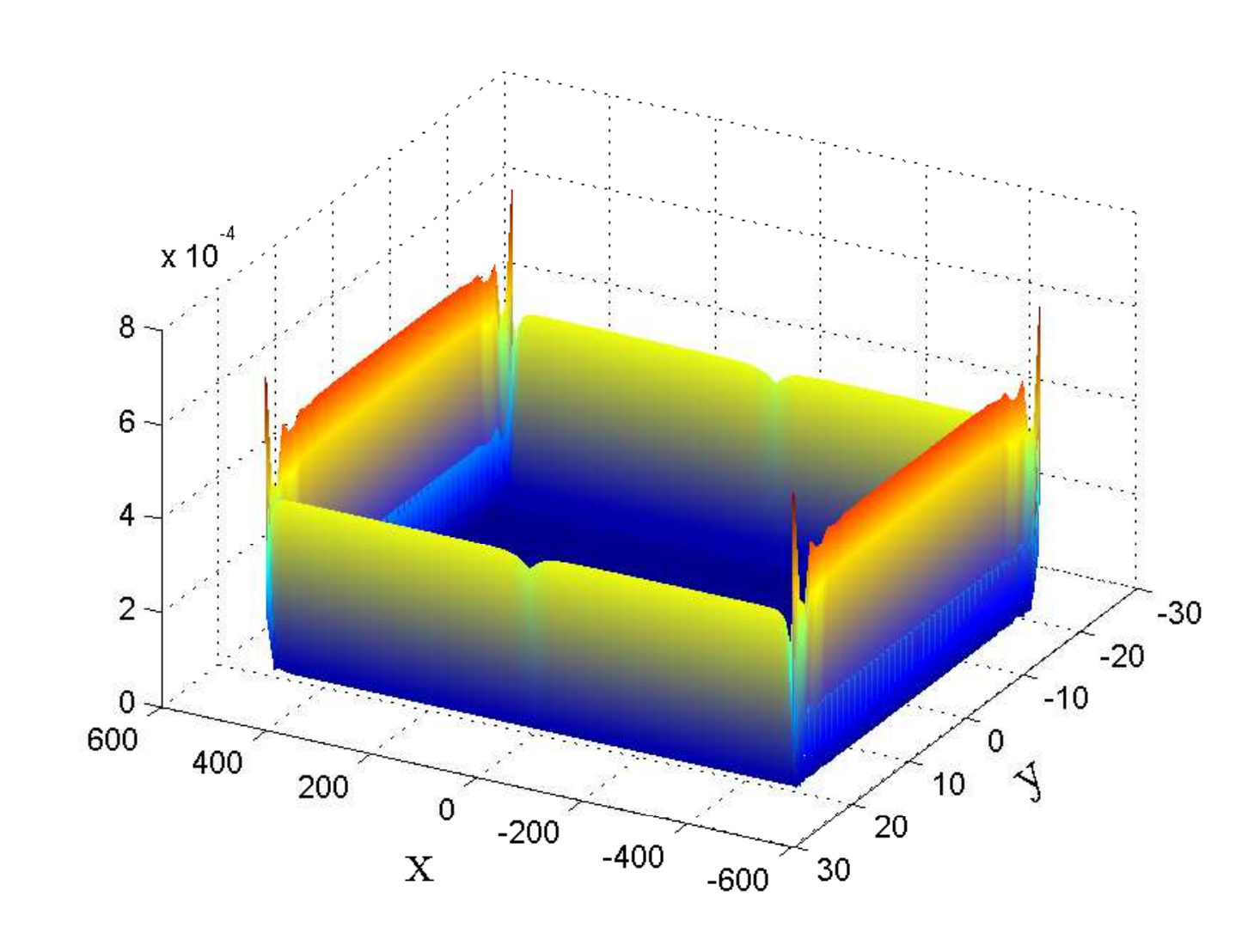}

(c)\includegraphics[scale=0.5]{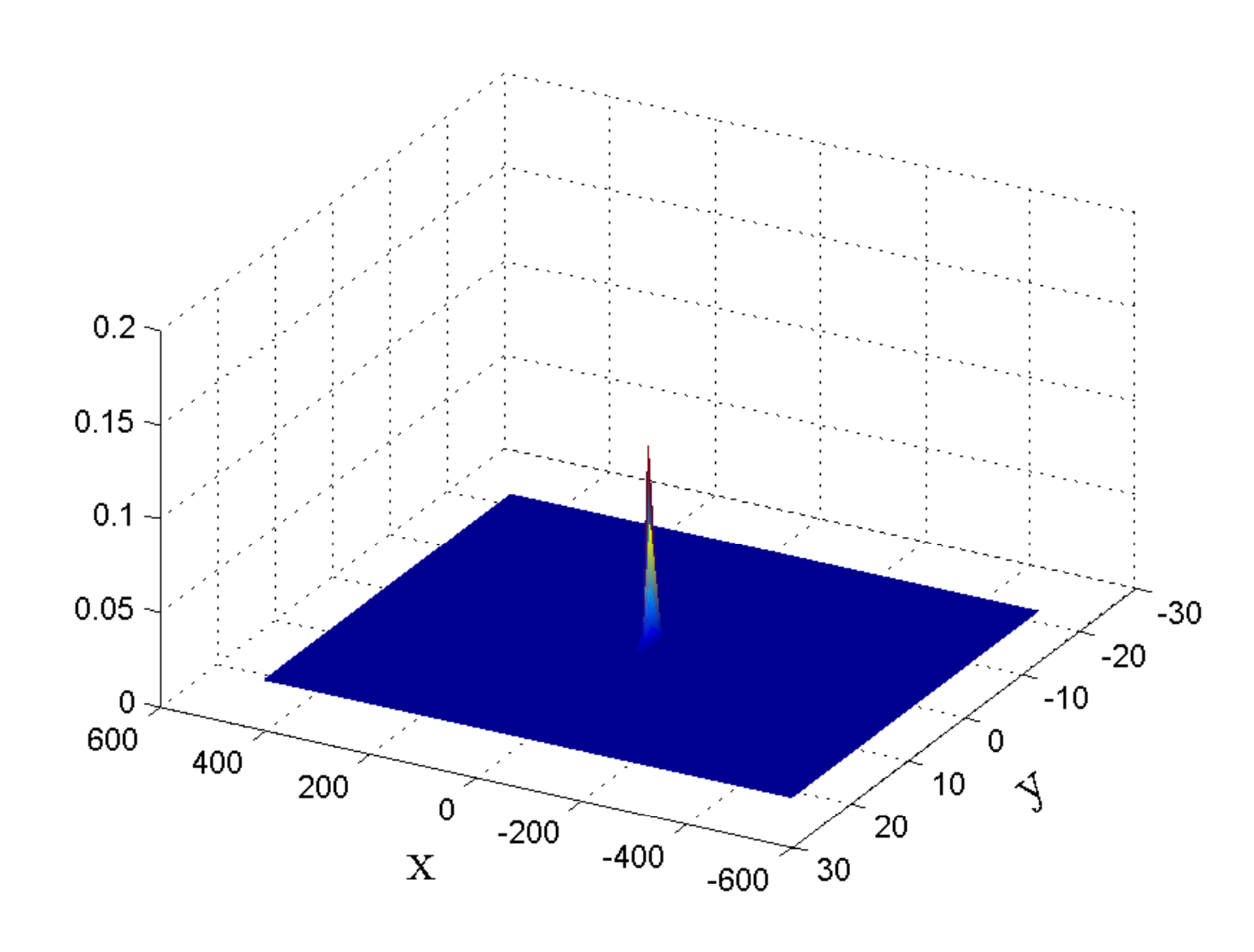}(d)\includegraphics[scale=0.5]{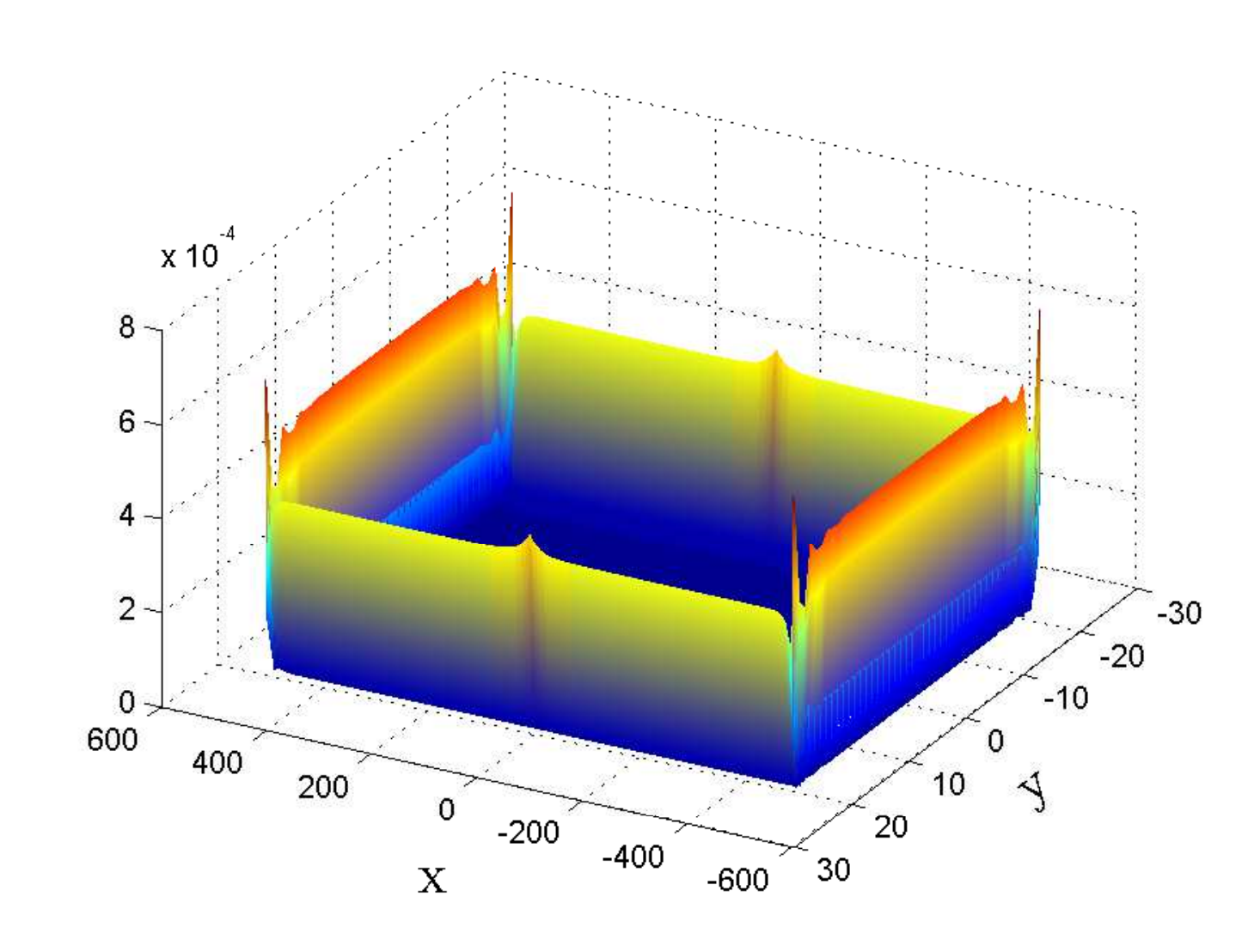}

\caption{The probability distribution of the Majorana zero modes in a system in the strong (1:00) phase, with a vortex at the center of the system. The coordinate along the wire is denoted by $x$, and the wire index by $y$. Panels (a) and (c) show the probability distribution of the zero mode at the vortex core; panels (b) and (d) show the distribution of the zero mode localized on the edge. In (a) and (b), the chiralities of the vortex and the superconductor are identical, whereas in (c) and (d) the chiralities are opposite. Notice the change in the distribution of the edge zero mode near the vortex (at $x=0$, $y=\pm 21$): in panel (b) the probability has a dip near the vortex, whereas in (d) it has a peak. This is because of the effect of the supercurrents around the vortex on the edge states. In case (a,b) the velocity of the edge mode increases in the vicinity of the vortex. Due to probability conservation, the probability current (given by the velocity times the probability) is divergence free. In order to compensate for the increase in velocity, the probability near the vortex must decreases. Following the same reasoning, in the case (c,d) (of opposite chirality), there is a peak in the probability distribution of the edge state near the vortex. In these calculations, we use the following parameters $t_{x}=1$, $t_{y}=0.3$, $\Delta=0.6$,
$\Delta_{y}=0$, $\mu=-1.4$, $V_{z}=0.6\sqrt{2}$, $\beta=0.3$,
and $\alpha=1$ [for which the ``sweet point'' conditions, Eq.(\ref{eq:"sweat point"-1}), are fulfilled].
\label{fig:The-probability-distribution with a vortex} }
\end{figure*}

We tested this effect numerically for a finite system with a single
vortex. Fig. \ref{fig:The-probability-distribution with a vortex}
shows the probability distribution of the two lowest energy states
in the system. We used two sets of parameters, one in which the vortex
has the same chirality, as shown in Fig. \ref{fig:The-probability-distribution with a vortex}a,
and the other in which it has opposite chirality Fig. \ref{fig:The-probability-distribution with a vortex}b,
relative to the superconductor. The low-energy states are superpositions
of a localized zero mode at the vortex core, and a propagating mode
localized on the edge. The variations of the probability distribution
of the edge mode are proportional to the variations of the local DOS
on the edge. As expected, the local DOS is either enhanced (second
row, Fig. \ref{fig:The-probability-distribution with a vortex}c and
Fig. \ref{fig:The-probability-distribution with a vortex}d) or suppressed
(first row, Fig. \ref{fig:The-probability-distribution with a vortex}a
and Fig. \ref{fig:The-probability-distribution with a vortex}b) in
the region of the edges close to the vortex. The vortex is incorporated
in our model by multiplying the order parameter by a spatially dependent
phase factor such that $\Delta(\vec{r})=\Delta e^{i\Phi(\vec{r})}$, where $e^{i\Phi(\vec{r})}=\frac{x+iy}{\sqrt{x^{2}+y^{2}}}$.
The vector potential was taken to be $\vec{A}=\frac{1}{2}B(x\hat{y}-y\hat{x})$.%
\footnote{We expect the splitting between the two eigenenergies closest to zero
to scale as $\exp(-L_{y}/\xi_{y})$, where $L_{y}$ is the size of
the system in the $y$ direction and $\xi_{y}$ is the corresponding
coherence length. (We assume that $L_{x}\gg\xi_{x}$.) The corresponding
eigenstates are linear combinations of the localized Majorana state
at the vortex core and a Majorana mode bound to the edge. The next
lowest energy scales as $\propto\frac{1}{L_{x}}$, due to the finite
size in the $x$ direction.%
}

The Doppler shift effect is a signature of all the topological phases.
It can be observed by measuring the tunneling current into the edge in the presence of
a supercurrent $J_{s}$ along the wire direction. Since, in the London
gauge, $J_{s}\propto A_{x}$, According to Eq.(\ref{eq:energy spectrum with orbital field-1}),
the low-energy DOS depends linearly on the current. 

\section{Distinguishing between the different phases experimentally }

\label{sec:Experimental-signature}In this section, we discuss ways to
distinguish experimentally between the different phases shown in Fig.
\ref{fig:phase-diagram-spinful}. The results are summarized in Table
\ref{tab:A-summery-of the phases}. These phases can be divided into
two groups: the strong topological phases characterized by $\nu\neq0$,
and the weak topological phases characterized by $\nu=0$. Each of
the phases is characterized by the appearance of Majorana modes
at specific momenta along the edge. The topological phases can be
determined by measuring the tunneling conductance from a metallic
lead into the edges.
\begin{table*}
\begin{centering}
{\small }%
\begin{tabular}{|>{\raggedright}p{0.3cm}|>{\centering}m{1.6cm}|>{\centering}m{2cm}|>{\centering}m{5cm}|>{\centering}m{1cm}|>{\centering}m{1cm}|>{\centering}p{1cm}|>{\centering}p{1cm}|}
\cline{1-6}
\multirow{2}{0.3cm}{} & \multirow{2}{1.6cm}[-0.2cm]{$\nu\mbox{\ensuremath{:}}\nu_{x,\pi}\nu_{y,\pi}${\footnotesize{} }} & \multirow{2}{2cm}[-0.2cm]{$\begin{array}{cc}
s_{(0,\pi)} & s_{(\pi,\pi)}\\
s_{(0,0)} & s_{(\pi,0)}
\end{array}$} & \multirow{2}{5cm}{{\footnotesize Edge states (non-chiral Majorana modes)}} & \multicolumn{2}{>{\centering}p{2cm}|}{{\footnotesize Tunneling DOS from a local tip }} & \multicolumn{1}{>{\centering}p{1cm}}{} & \multicolumn{1}{>{\centering}p{1cm}}{}\tabularnewline
\cline{5-6}
 &  &  &  & {\footnotesize $x$} & {\footnotesize $y$} & \multicolumn{1}{>{\centering}p{1cm}}{} & \multicolumn{1}{>{\centering}p{1cm}}{}\tabularnewline
\cline{1-6}
\multicolumn{1}{|>{\centering}m{0.3cm}|}{\begin{sideways}
Trivial
\end{sideways}} & {\footnotesize $0\mbox{:}00$} & {\footnotesize $\begin{array}{cc}
+ & +\\
+ & +
\end{array}$} & {\footnotesize None} & {\footnotesize 0} & {\footnotesize 0} & \multicolumn{1}{>{\centering}p{1cm}}{} & \multicolumn{1}{>{\centering}p{1cm}}{}\tabularnewline
\cline{1-6}
\multirow{3}{0.3cm}[-0.3cm]{\begin{sideways}
weak phases
\end{sideways}} & {\footnotesize $0\mbox{:}01$} & {\footnotesize $\begin{array}{cc}
- & +\\
- & +
\end{array}$} & {\footnotesize Along the edge $\parallel y$ } & {\footnotesize 0} & {\footnotesize Finite} & \multicolumn{1}{>{\centering}p{1cm}}{} & \multicolumn{1}{>{\centering}p{1cm}}{}\tabularnewline
\cline{2-6}
 & {\footnotesize $0\mbox{:}10$} & {\footnotesize $\begin{array}{cc}
+ & +\\
- & -
\end{array}$} & {\footnotesize Along the edge $\parallel x$} & {\footnotesize Finite} & {\footnotesize 0} & \multicolumn{1}{>{\centering}p{1cm}}{} & \multicolumn{1}{>{\centering}p{1cm}}{}\tabularnewline
\cline{2-6}
 & {\footnotesize $0\mbox{:}11$} & {\footnotesize $\begin{array}{cc}
- & +\\
+ & -
\end{array}$} & {\footnotesize Along the edges $\parallel x$ and $y$} & {\footnotesize Finite} & {\footnotesize Finite} & \multicolumn{1}{>{\centering}p{1cm}}{} & \multicolumn{1}{>{\centering}p{1cm}}{}\tabularnewline
\cline{1-6}
\multicolumn{1}{>{\raggedright}p{0.3cm}}{} & \multicolumn{1}{>{\centering}m{1.6cm}}{} & \multicolumn{1}{>{\centering}m{2cm}}{} & \multicolumn{1}{>{\centering}m{5cm}}{} & \multicolumn{1}{>{\centering}m{1cm}}{} & \multicolumn{1}{>{\centering}m{1cm}}{} & \multicolumn{1}{>{\centering}p{1cm}}{} & \multicolumn{1}{>{\centering}p{1cm}}{}\tabularnewline
\hline
\multirow{3}{0.3cm}{} & \multirow{3}{1.6cm}{$\nu\mbox{\ensuremath{:}}\nu_{x,\pi}\nu_{y,\pi}${\footnotesize{} }} & \multirow{3}{2cm}{$\begin{array}{cc}
s_{(0,\pi)} & s_{(\pi,\pi)}\\
s_{(0,0)} & s_{(\pi,0)}
\end{array}$} & \multirow{3}{5cm}{{\footnotesize Chiral Edge states along the edges $\parallel x$ and
$y$, located near $k_{x}=k_{x,0}$ and}\\
{\footnotesize{} $k_{y}=k_{y,0}$, respectively}} & \multicolumn{4}{c|}{{\footnotesize Tunneling DOS from a wire }}\tabularnewline
\cline{5-8}
 &  &  &  & {\footnotesize $k_{y}=0$} & {\footnotesize $k_{x}=0$} & $k_{y}=\pi$ & $k_{x}=\pi$\tabularnewline
 &  &  &  &  &  &  & \tabularnewline
\hline
\multirow{5}{0.3cm}[-0.7cm]{\begin{sideways}
Strong phases
\end{sideways}} & {\footnotesize $1\mbox{:}00$} & {\footnotesize $\begin{array}{cc}
+ & +\\
- & +
\end{array}$} & {\footnotesize $k_{x,0}=0$, $k_{y,0}=0$} & {\footnotesize Finite} & {\footnotesize Finite} & 0 & 0\tabularnewline
\cline{2-8}
 & {\footnotesize $1\mbox{:}01$} & {\footnotesize $\begin{array}{cc}
- & +\\
+ & +
\end{array}$} & {\footnotesize $k_{x,0}=0$, $k_{y,0}=\pi$} & {\footnotesize Finite} & {\footnotesize 0} & 0 & {\footnotesize Finite}\tabularnewline
\cline{2-8}
 & {\footnotesize $1\mbox{:}10$} & {\footnotesize $\begin{array}{cc}
+ & +\\
+ & -
\end{array}$} & {\footnotesize $k_{x,0}=\pi$, $k_{y,0}=0$} & {\footnotesize 0} & {\footnotesize Finite} & {\footnotesize Finite} & 0\tabularnewline
\cline{2-8}
 & {\footnotesize $1\mbox{:}11$} & {\footnotesize $\begin{array}{cc}
+ & -\\
+ & +
\end{array}$} & {\footnotesize $k_{x,0}=\pi$, $k_{y,0}=\pi$} & {\footnotesize 0} & {\footnotesize 0} & {\footnotesize Finite} & {\footnotesize Finite}\tabularnewline
\cline{2-8}
 & {\footnotesize $2\mbox{:}11${*}} & {\footnotesize $\begin{array}{cc}
- & +\\
+ & -
\end{array}$} & {\footnotesize Two edge modes at $k_{x,0}=0$, $k_{y,0}=0$}\\
{\footnotesize and $k_{x,0}=\pi$, $k_{y,0}=\pi$} & {\footnotesize Finite} & {\footnotesize Finite} & {\footnotesize Finite} & {\footnotesize Finite}\tabularnewline
\hline
\end{tabular}
\par\end{centering}{\small \par}

\caption{\label{tab:A-summery-of the phases}Summary of the different phases,
their edge properties, and their experimental signatures. The different
phases are labeled by the three indices $\nu\mbox{\ensuremath{:}}\nu_{x,\pi}\nu_{y,\pi}$,
where $\nu\in\mathbb{Z}$ and $\nu_{x,\pi},\nu_{y,\pi}\in\mathbb{Z}_{2}$.
The strong index $\nu$ is given by Eq.(\ref{eq:Chern}), and the
weak indices, $\nu_{x,\pi}$ and $\nu_{y,\pi}$, are defined by the
product of a pair of $s_{\mathbf{\boldsymbol{\Gamma}}_{i}}$'s {[}see
Eq.(\ref{eq:nu}){]}. The values of $s_{\mathbf{\Gamma}_{i}}$ in
each phase are listed in the second column. The third column describes
the edge states. The edge properties can be determined from the $s_{\mathbf{\Gamma}_{i}}$'s.
For example, a low-energy state appears along the edge parallel to
$x$ near momentum $k_{x}=0$ if $s_{(0,0)}s_{(0,\pi)}=-1$, and so
forth.\protect \\
In the upper table, the first line is the trivial phase. Lines 2-4
describe the weak topological phases, which can be distinguished by
measuring the local tunneling density of states at low energy on the
edges perpendicular and parallel to the wires.\protect \\
The lower table lists the strong topological phases. These phases
cannot be distinguished from each other using a local tunneling probe,
since they all have low-energy chiral modes on all edges. The phases
differ in the position of the edge states in momentum space (see Fig.
\ref{fig:spectrum in v-0 and v=00003D2}). Hence, they can be distinguished
in a tunneling experiment from an extended wire, such that the momentum
along the edge is conserved in the tunneling process (as in the experiments
of Ref. \onlinecite{auslaender2004many}). The momentum probed in
the tunneling experiment can be tuned by varying the flux between
the wire and the edge of the system\cite{auslaender2004many}. \protect \\
({*}) The $0\mbox{:}11$ and $2\mbox{:}11$ phases have identical
weak indices. Therefore, they cannot be distinguished by tunneling
experiments. In order to resolve them, one would need additional experiments
(e.g., a measurement of the thermal Hall conductance\cite{read2000paired}).}
\end{table*}

The strong phases are characterized by the presence of chiral edge
states along any edge, regardless of its direction. The weak phases,
on the other hand, may have low-energy edge states on a boundary along
$x$, $y$, or both, depending on the values of the weak indices. The presence of an edge state near a given momentum can be determined from the indices $s_{\mathbf{\Gamma}_i}$ [defined above Eq.~(\ref{eq:nu})]. 
For example, a low-energy state appears along the edge parallel to
$x$ near momentum $k_{x}=0$ if $s_{(0,0)}s_{(0,\pi)}=-1$, and so
forth.

Scanning tunneling spectroscopy (STS) experiments, in which electrons
tunnel from a point-like tip into the sample, will detect a finite
tunneling conductance at low energy on all boundaries in the strong
phase. Finding a finite conductance on a boundary along $x$ but zero
conductance on a boundary along $y$ (or vise versa) is a signature
of a weak phase. In all cases, the bulk of the system is fully gapped.

Phases with different weak indices can be distinguished in momentum-resolved
tunneling experiments. One can imagine tunneling from an extended
wire into the boundary of the system, so that the momentum along the
boundary is approximately conserved. A perpendicular magnetic field
can be used to control the momentum transfer in the tunneling process\cite{auslaender2004many}.
This way, one can isolate the contributions to the tunneling density
of states from the vicinity of $k=0$ and $k=\pi$ along the boundary.

Table \ref{tab:A-summery-of the phases} summarizes the different
possible phases and their signatures in STS experiments, as well as
momentum-resolved tunneling experiments at momentum $k=0$ or $\pi$
along the boundary. By combining these experiments, many of the phases
can be distinguished from each other. However, the $0\mbox{:}11$,
and $2\mbox{:}11$ phases have identical weak indices (see Table \ref{tab:A-summery-of the phases}),
hence tunneling experiments can not distinguish between them. In order
to resolve them, one would need additional experiments (e.g. a measurement
of the thermal Hall conductance\cite{read2000paired}). Notice, however,
that neither the 0:11 nor the $2\mbox{:}11$ phases occur naturally
in our coupled-wire setup (see dashed region in Fig. \ref{fig:phase-diagram-spinful}).
As discussed in subsection \ref{sub:-phase v=00003D2 } the $2\mbox{:}11$
phase can be realized in a super-lattice structure, and appendix \ref{sec:The-topological-phase 011}
describes a possible realization of the $0\mbox{:}11$ phase.

\section{Conclusions}

\label{sec: conclusions}In this work, we studied an array of weakly
coupled superconducting wires with spin orbit coupling. This system
can be used to realize a rich variety of two-dimensional topological
phases, either of the ``weak'' or ``strong'' kind. One can tune between
different phases using experimentally accessible parameters, such
as the chemical potential and the magnetic field. The strong phases
are anisotropic analogous to the chiral \textit{$p+ip$} phase, and
have chiral (Majorana) edge modes at their boundaries.

In particular, there is a choice of parameters such that the chiral
edge states are almost completely localized on the two outmost wires.
At this ``sweet point'', the edge states on the two opposite edges
do not mix even in a system with few wires. Similarly, at this point
in parameters space, the Majorana zero mode at a vortex core is tightly
localized in the direction perpendicular to the wires, and resides
only on one or two wires.

Each one of the topological phases has a unique signature in its edge
spectrum. The different phases can be distinguished in tunneling experiments
into the edge. Moreover, density of states of the Majorana edge modes
is predicted to vary linearly with an applied supercurrent in the
wires.
\begin{acknowledgments}
We would like to acknowledge discussions with Y. E. Kraus, A. Keselman,
A. Haim, Y. Schattner, and Y. Werman. E. B. was supported by the Israel
Science Foundation, by a Marie Curie CIG grant, by the Minerva foundation, by a German-Israeli Foundation (GIF) grant, and by the Robert
Rees Fund. Y.O. was supported by an Israel Science Foundation grant,
by the ERC advanced grant, by the Minerva foundation, by the TAMU-WIS
grant, and by a DFG grant.
\end{acknowledgments}
\appendix
 \textbf{}

\section{\label{sec:The-topological-phase 011}The topological phase 0:11 }

\begin{figure}
\includegraphics[bb=0bp 0bp 316bp 316bp,scale=0.5]{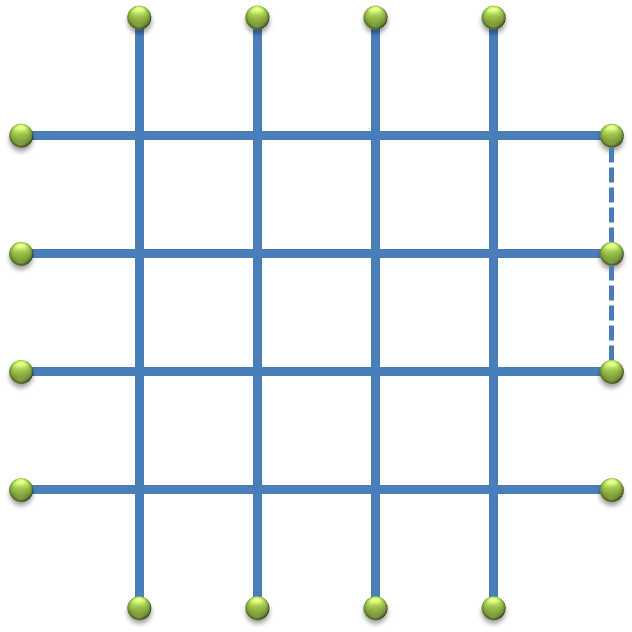}

\caption{A schematic illustration of a system in the 0:11 phase. The system is composed of two layers of weakly coupled wires,
rotated by $90^{o}$ with respect to each other. Each wire has a Majorana zero mode at its ends; these zero modes form a low-energy band when a coupling between the wires is introduced. \label{fig: Phase 0:11-1} }
\end{figure}

The topological phase $0\mbox{:}11$ does not appear in the physical
system of coupled wires, or in the simplified model described in Sec.
\ref{sec: Overview  Topological  supercondu}. One can imagine a
different setup that realizes this phase, such as the one presented
in Fig. \ref{fig: Phase 0:11-1}. Consider two layers of weakly coupled
parallel wires, rotated by $90^{o}$ relative to each other. Initially,
suppose that there is no coupling between the two layers. One of the
layers is in the $0\mbox{:}01$ phase (i.e., adiabatically connected
to a phase of decoupled 1D topological superconductors oriented along
the $x$ direction), and the other is in the $0\mbox{:}10$ phase.
If we turn on a weak coupling between the two layers, we get a single
two-dimensional system whose indices are the sums of the indices of
the two constituents (modulo 2). The entire system is therefore in
the $0\mbox{:}11$ phase.

The $0\mbox{:}11$ phase is characterized by having non-chiral Majorana
edge modes at low energy on edges along the $x$ or $y$ directions.
More generally, on an edge directed along an angle $\theta$ such
that $\tan\theta=p/q$ (where $p,q$ are integers), there is a low
energy Majorana mode if $p+q$ is odd. This can be seen from fact
that such an edge preserves translational invariance along $(p,q)$.
If we turn off the coupling between all the wires (both within each
layer and between the layers), every unit cell of the edge contains
$p+q$ Majorana zero modes. Upon turning on inter-wire coupling, we
get an edge mode that crosses zero energy at momenta $0$ and $\pi$
parallel to the edge if the number of Majorana zero modes per primitive
unit cell is odd. Otherwise, we can pair the Majorana zero modes within
each unit cell and gap them out.

It is interesting to consider the case of an edge such that $\tan\theta$
is irrational. In this case, the edge breaks translational symmetry,
so we cannot define the number of Majorana modes in a unit cell in
the decoupled limit. On physical grounds, we expect to get a gapless
mode in this case, since we can approximate $\tan\theta$ by $p/q$
arbitrarily well, with $p+q=$odd.

\section{\label{Appendix D}Computation of the topological invariants in an
array of coupled quantum wires}

The Hamiltonian Eq.(\ref{eq:Hwires}) in momentum space is given by
$\mathcal{H}=\frac{1}{2}\sum\psi_{\mathbf{k}}^{\dagger}h(\mathbf{k})\psi_{\mathbf{k}}$,
where $\mbox{\ensuremath{\mathbf{k}}}=\left(k_{x},k_{y}\right)$,
and
\begin{multline}
h(\mathbf{k})=\xi_{\mathbf{k}}\tau_{z}+\alpha\mbox{sin}(k_{x})\tau_{z}\sigma_{y}+V_{z}\sigma_{z}+\\
\left[\Delta+\Delta_{y}\mbox{cos}(k_{y})\right]\tau_{x}+\beta\mbox{sin}(k_{y})\tau_{z}\sigma_{x}.\label{eq:H(kx,ky)}
\end{multline}
Here, $\xi_{\mathbf{k}}=-2t_{x}\mbox{cos}(k_{x})-2t_{y}\mbox{cos}(k_{y})-\mu$.
The Hamiltonian $h(\mathbf{k})$ has particle-hole symmetry, $Ch(\mathbf{k})C^{-1}=-h(-\mathbf{k})$,
where $C=\tau_{y}\sigma_{y}\mathcal{K}$. According to Eqs. (\ref{eq:nu}),
the topological indices depend only on the Hamiltonian at the high
symmetry points $\mathbf{\boldsymbol{\Gamma}}_{i}$. At these points,
it is convenient to transform to the Majorana basis: $\psi_{\mathbf{\boldsymbol{\Gamma}}_{i}}=V\gamma_{\mathbf{\boldsymbol{\Gamma}}_{i}}$,
where $\gamma_{\mathbf{\boldsymbol{\Gamma}}_{i}}^{T}=\left(\begin{array}{cccc}
\chi_{1} & \chi_{2} & \chi_{3} & \chi_{4}\end{array}\right)$ ($\chi_{1\dots4}$ are Majorana operators) and

\begin{equation}
V=\frac{1}{2}\left(\begin{array}{cccc}
1 & 0 & i & 0\\
0 & 1 & 0 & i\\
0 & 1 & 0 & -i\\
-1 & 0 & i & 0
\end{array}\right).
\end{equation}
In this basis, the Hamiltonians at the high symmetry points are given
by
\begin{multline}
B_{\mathbf{\boldsymbol{\Gamma}}_{i}}=V^{\dagger}H_{\mathbf{\boldsymbol{\Gamma}}_{i}}V=\frac{1}{2}\tau_{y}(\xi_{\mathbf{\boldsymbol{\Gamma}}_{i}}+V_{z}\sigma_{z}-i\Delta\sigma_{y})=\\
\frac{i}{2}\left(\begin{array}{cccc}
0 & 0 & \xi_{\mathbf{\boldsymbol{\Gamma}}_{i}}+V_{z} & -\Delta\\
0 & 0 & \Delta & \xi_{\mathbf{\boldsymbol{\Gamma}}_{i}}-V_{z}\\
-(\xi_{\mathbf{\boldsymbol{\Gamma}}_{i}}+V_{z}) & -\Delta & 0 & 0\\
\Delta & -(\xi_{\mathbf{\boldsymbol{\Gamma}}_{i}}-V_{z}) & 0 & 0
\end{array}\right).
\end{multline}
The Pfaffians of $B_{\mathbf{\boldsymbol{\Gamma}}_{i}}$ can be computed
using the relation
\[
\mathrm{Pf}\left(\begin{array}{cccc}
0 & a & b & c\\
-a & 0 & d & e\\
-b & -d & 0 & f\\
-c & -e & -f & 0
\end{array}\right)=af-be+dc.
\]
In our case, $c=-\Delta$, $d=\Delta$, $a=0$, $d=\xi_{\mathbf{\boldsymbol{\Gamma}}_{i}}+V_{z}$,
and $e=\xi_{\mathbf{\boldsymbol{\Gamma}}_{i}}-V_{z}$. This gives
$s_{\mathbf{\boldsymbol{\Gamma}}_{i}}={\rm sign}\left\{ i\left[\mbox{Pf}(B_{\mathbf{\boldsymbol{\Gamma}}_{i}})\right]\right\} ={\rm sign}\left(\xi_{\mathbf{\boldsymbol{\Gamma}}_{i}}^{2}-V_{z}^{2}+\Delta^{2}\right)$.
From this, the $\mathbb{Z}_{2}$ indices can be computed using Eq.(\ref{eq:nu}).
It also enables us to produce the phase diagram as depicted in Fig.
\ref{fig:phase-diagram-spinful} up to the parity of the Chern number
using Eq.(\ref{eq:weak and strong relation}). We have calculated
the Chern number numerically and verified these conclusions. The calculation
was done with periodic boundary conditions in both directions, using
the formula\cite{bernevig2013topological}:
\begin{equation}
\nu=\frac{1}{\pi}\underset{\mathbf{k}\in BZ}{\sum}\underset{\alpha\beta\in x,y}{\sum}\varepsilon_{\alpha\beta}\mathrm{Tr}(P(\mathbf{k})\partial_{\alpha}P(\mathbf{k})\partial_{\beta}P(\mathbf{k})),\label{eq:CN using P}
\end{equation}
where $P(\mathbf{k})$ is a projection operator on the negative energy
bands, defined as $P(\mathbf{k})=\underset{\{E_{n}<0\}}{\sum}\vert\psi_{n}\rangle\langle\psi_{n}\vert$.
Substituting $P(\mathbf{k})$ into Eq.(\ref{eq:CN using P}), we arrive
at Eq.(\ref{eq:Chern}). In Appendix \ref{Appendix C} we give an
analytical explanation for the existence of the strong index $\nu=2$
in the phase diagram.

\section{\label{Appendix C}Explanation for the existence of the strong index
$\nu=2$ in the phase diagram }

In order to understand the appearance of a phase with a strong index $\nu=2$
in the model described in Sec.~\ref{sec:spinfull model}, we analyze
the changes in the Chern number along a trajectory in the parameter
space $V_{z}-\mu$. Along the line $\mu=0$ and $-\infty<V_{z}<\infty$,
the gap closes at $V_{z}=\pm\sqrt{\Delta_{\mathrm{eff}}^{2}+(2t_{x}\pm2t_{y})^{2}}$
{[}see Eq.(\ref{eq:spectrum of the system}){]}. At each of these
points in the parameter space, the closure of the gap occurs simultaneously
at two points in the Brillouin zone. E.g., at $V_{z}=\sqrt{\Delta_{\mathrm{eff}}^{2}+(2t_{x}+2t_{y})^{2}}=V_{z}^{\mathrm{c_{+}}}$,
the gap closes at $\mathbf{k}=(0,0)$ and $(\pi,\pi)$, whereas
at $V_{z}=\sqrt{\Delta_{\mathrm{eff}}^{2}+(2t_{x}-2t_{y})^{2}}\equiv V_{z}^{\mathrm{c_{-}}}$,
the gap closes at $\mathbf{k}=(\pi,0)$ and $(0,\pi)$.

 Near these
points, the low-energy part of the spectrum is linear (Dirac-like).
Tuning $V_{z}$ away from $V_{z}^{\mathrm{c_{\pm}}}$, Dirac mass
terms appear. The vicinity of each Dirac point in the Brillouin zone
contributes $\mathrm{sign}(m)/2$ to the total Chern number, where
$m$ is the Dirac mass. Thus, a sign change in the mass term corresponds
to a change in the total Chern number by $\pm1$. Below, we derive
the form of the massive Dirac Hamiltonians at the vicinity of $V_{z}=V_{z}^{\mathrm{c_{+}}}$. We show
that the Chern number changes by $\pm2$ as $V_{z}$ is swept through
$V_{z}^{\mathrm{c_{+}}}$.

First, we project the Hamiltonian in Eq.~(\ref{eq:H(kx,ky)}) to the two
bands closest to zero energy. At the critical point $V_{z}=V_{z}^{\mathrm{c_{+}}}$,
the zero-energy eigenstates are $v_{1,\mathbf{\boldsymbol{\Gamma}}_{i}}^{T}=(\begin{array}{cccc}
-1 & 0 & \frac{\xi_{\mathbf{\boldsymbol{\Gamma}}_{i}}+V_{z}}{\Delta} & 0\end{array})$, $v_{2,\mathbf{\boldsymbol{\Gamma}}_{i}}^{T}=\left(\begin{array}{cccc}
0 & \frac{\xi_{\mathbf{\boldsymbol{\Gamma}}_{i}}+V_{z}}{\Delta} & 0 & 1\end{array}\right)$. \textcolor{black}{Here $\boldsymbol{\Gamma}_{1}=(0,0)$, $\boldsymbol{\Gamma}_{2}=(\pi,\pi)$,
and} $\xi_{\mathbf{k}}=-\mu-2t_{x}\cos(k_{x})-2t_{y}\cos(k_{y})$\textcolor{black}{.}
The effective $2\times2$ Hamiltonians in the vicinity of $\boldsymbol{\Gamma}_{1,2}$
and near the critical point $V_{z}^{\mathrm{c}_{+}}$ are given by
\begin{multline*}
H_{2\times2}(\boldsymbol{\Gamma}_{1})=\frac{V_{z}^{2}-V_{z,c_{+}}^{\mathrm{2}}}{2V_{z}}\sigma_{z}-\frac{\alpha\Delta}{V_{z}}\delta k_{x}\sigma_{y}-\frac{\beta\Delta}{V_{z}}\delta k_{y}\sigma_{x},
\end{multline*}

\begin{multline*}
H_{2\times2}(\boldsymbol{\Gamma}_{2})=\frac{V_{z}^{2}-V_{z,c_{+}}^{\mathrm{2}}}{2V_{z}}\sigma_{z}+\frac{\alpha\Delta}{V_{z}}\delta k_{x}\sigma_{y}+\frac{\beta\Delta}{V_{z}}\delta k_{y}\sigma_{x}.
\end{multline*}
 Here, $\delta\mathbf{k}=\mathbf{k}-\mathbf{\boldsymbol{\Gamma}}_{i}$.

\begin{flushleft}
At sufficiently large $V_{z}$, the strong index is $\nu=0$. Crossing
through $V_{z}=V_{z}^{\mathrm{c_{+}}}$, the contributions to the
Chern number from the vicinity of $\mathbf{\boldsymbol{\Gamma}}_{i}$ is given by
\begin{multline}
\nu_{\boldsymbol{\Gamma}_{1}}=\nu_{\boldsymbol{\Gamma}_{2}}=\frac{1}{2}\mbox{sign}(\frac{V_{z}^{2}-V_{z,c_{+}}^{\mathrm{2}}}{2V_{z}})\mbox{sign}(\frac{\Delta^{2}}{V_{z}^{2}}\alpha\beta).
\end{multline}

\par\end{flushleft}

Therefore, the change in the Chern number from $V_{z}>V_{z}^{\mathrm{c_{+}}}$
to $V_{z}<V_{z}^{\mathrm{c_{+}}}$ is $\Delta\nu=\Delta\nu_{\boldsymbol{\Gamma}_{1}}+\Delta\nu_{\boldsymbol{\Gamma}_{2}}=-2$.
In the same way, tuning $V_{z}$ through $V_{z}^{\mathrm{c_{-}}}$
from above changes the Chern number by $\Delta\nu=2$.

\section{Quasi-particle spectrum in the presence of orbital magnetic field }

\label{sec:appendix A the spectrum with an orbital field }In this
appendix, we will show that the density of states of the edge modes
in the presence of an orbital magnetic field (or a supercurrent parallel
to the edge) is linear in the vector potential. We consider a BdG
Hamiltonian of the form $\mathcal{H}=\underset{\mathbf{k}}{\frac{1}{2}\sum}\Psi_{\mathbf{k}}^{\dagger}h(\mathbf{k},\mathbf{A})\Psi_{\mathbf{k}}$,
using the Nambu notation: $\Psi_{\mathbf{k}}^{\dagger}=\left(\psi_{\uparrow,\mathbf{k}}^{\dagger},\psi_{\downarrow,\mathbf{k}}^{\dagger}\psi_{\downarrow,-\mathbf{k}},-\psi_{\uparrow,-\mathbf{k}}\right)$,
and

\begin{equation}
h(\mathbf{k},\mathbf{A})=\left(\begin{array}{cc}
H_{0}(\mathbf{k},\mathbf{A}) & \Delta_{\mathbf{k}}\\
\Delta_{\mathbf{k}}^{\dagger} & -\sigma_{y}H_{0}^{*}(\mathbf{k},-\mathbf{A})\sigma_{y}
\end{array}\right).
\end{equation}
Here, $H_{0}(\mathbf{k},\mathbf{A})$ is the normal (non-superconducting)
Hamiltonian, and $\Delta_{\mathbf{k}}$ is the pairing matrix. We
note that the particle-hole symmetry is expressed in this basis as
follows: $\mathcal{C}=\tau_{y}\sigma_{y}K$. This implies $h(\mathbf{k},\mathbf{A})=-\mathcal{C}h(\mathbf{k},\mathbf{A})\mathcal{C}$.
Hence, the spectrum should satisfy
\begin{equation}
E(\mathbf{k},\mathbf{A})=-E(-\mathbf{k},\mathbf{A})\label{eq:PHS to E}
\end{equation}
.

We now consider a semi-infinite system with periodic boundary conditions
in the $x$ direction and half-infinite in the $y$ direction. The
equations for the eigenstates acquire the form: $h(k_{x},\mathbf{A})\vert\psi_{m}(k_{x},\mathbf{A})\rangle=E_{m}(k_{x},\mathbf{A})\vert\psi_{m}(k_{x},\mathbf{A})\rangle$,
where $h(k_{x},\mathbf{A})$ is a matrix whose indices are the wire
labels, and $m$ runs over all the eigenstates.

Let us assume that there is a single Majorana edge mode at $k_{x}=0$
and that $\mathbf{A}=A_{x}\hat{x}$. Expanding the edge mode energy,
$E_{0}(k_{x},A_{x})$, near $k_{x}=0$ and $A_{x}=0$, we get
\begin{multline*}
E_{0}(k_{x},A_{x})=E_{0}(0,0)+k_{x}\partial_{k_{x}}E_{0}|_{k_{x},A_{x}=0}+\\
A_{x}\partial_{A_{x}}E_{0}|_{k_{x},A_{x}=0}+k_{x}A_{x}\partial_{k_{x}}\partial_{A_{x}}E_{0}|_{k_{x},A_{x}=0}+\cdots.
\end{multline*}
Particle-hole symmetry {[}see Eq.(\ref{eq:PHS to E}){]} requires
that $E_{0}(0,0)=0$ and $\partial_{A_{x}}E_{0}|_{k_{x},A_{x}=0}=0$.
We denote $v_{x}=\partial_{k_{x}}E_{0}|_{k_{x},A_{x}=0}$, $\delta_{x}=\partial_{k_{x}}\partial_{A_{x}}E_{0}|_{k_{x},A_{x}=0}$.
The coefficients $v_{x}$ and $\delta_{x}$ can be calculated using:

\begin{equation}
v_{x}=\langle\varphi_{0}|\partial_{k_{x}}h|\varphi_{0}\rangle,\label{eq:vx}
\end{equation}

\begin{align}
\delta_{x} & =\partial_{A_{x}}\langle\varphi_{0}\vert\partial_{k_{x}}h\vert\varphi_{0}\rangle\nonumber \\
 & =\left(\langle\varphi_{0}\vert\partial_{A_{x}}\partial_{k_{x}}h\vert\varphi_{0}\rangle+\langle\partial_{A_{x}}\varphi_{0}\vert\partial_{k_{x}}h\vert\varphi_{0}\rangle+\mathrm{c.c.}\right).\label{eq:deltax}
\end{align}
Substituting $\vert\partial_{A_{x}}\varphi_{0}\rangle=\underset{m\neq0}{\sum}\frac{\vert\varphi_{m}\rangle\langle\varphi_{m}\vert\partial_{A_{x}}h\vert\varphi_{0}\rangle}{E_{m}-E_{0}}$,
 we can arrive to the following formula:

\begin{multline}
\langle\partial_{A_{x}}\varphi_{0}\vert\partial_{k_{x}}h\vert\varphi_{0}\rangle=\underset{m\neq0}{\sum}\frac{\langle\varphi_{0}\vert\partial_{k_{x}}h\vert\varphi_{m}\rangle\langle\varphi_{m}\vert\partial_{A_{x}}h\vert\varphi_{0}\rangle}{E_{m}-E_{0}}.\label{eq:2nd}
\end{multline}
In Eqs.(\ref{eq:vx}), (\ref{eq:deltax}), and (\ref{eq:2nd}), $\vert\varphi_{m}\rangle$
is evaluated at $k_{x}=A_{x}=0$.

For the model described in Sec. \ref{sec:spinfull model}, $\Delta_{\mathbf{k}}=I_{2\times2}\Delta$
and $H_{0}$ has the following form near $k_{x}=0$:

\begin{multline*}
H_{0}(\mathbf{k},A)=t_{x}(k_{x}-A_{x})^{2}-2t_{y}\cos(k_{y})-\mu\\
-2t_{x}+\alpha k_{x}\sigma_{y}+\beta k_{y}\sigma_{x}+V_{z}\sigma_{z}.
\end{multline*}
In the sweet point, discussed in \ref{sub:``Sweet-point''}, the edge
state at $k_{x}=0$ is completely localized on the outmost wire.
In the limit $V_{z}\gg t_{y},\beta$, we can evaluate Eqs.(\ref{eq:vx}),
(\ref{eq:deltax}), and (\ref{eq:2nd}) perturbatively in $t_{y}/V_{z}$
and $\beta/V_{z}$. To zeroth order, we can replace the wave-functions
$\vert\varphi_{m}\rangle$\textcolor{red}{{} }($m\ne0$) with the eigenstates
of the decoupled wires ($t_{y}=\beta=0$). Using the explicit form
of the single wire eigenstates, $\psi_{1,k_{x},j}$, $\psi_{2,k_{x},j}$
and their particle-hole partners {[}given in Eqs.(\ref{eq:W}) and
(\ref{eq:Psi}){]}, we get

\begin{equation}
v_{x}=\alpha\langle\psi_{1}\vert\tau_{z}\sigma_{y}\vert\psi_{1}\rangle=\frac{\Delta\alpha}{V_{z}},
\end{equation}

\begin{align*}
\delta_{x} & =t_{x}.
\end{align*}

\section{The effect of the orbital field on the ``sweet point'' \label{sec:the orbital field and the =00201Csweet point=00201D }}

Here, we demonstrate that the ''sweet point'' condition {[}Eq.(\ref{eq:"sweat point"-1}){]}
is not modified in the presence of a small orbital magnetic field.

Let us consider a system with a uniform orbital field. The single-wire
part of the Hamiltonian is given by:
\begin{multline}
\mathcal{H}_{j}=\int dk_{x}\varepsilon_{j}(k_{x},\phi)\psi_{k_{x},j}^{\dagger}\psi_{k_{x},j}\\
-\alpha\mbox{sin}\left(k_{x}-2\pi\phi j\right)\psi_{k_{x},j}^{\dagger}\sigma_{y}\psi_{k_{x},j}+\\
+V_{z}\psi_{k_{x},j}^{\dagger}\sigma_{z}\psi_{k_{x},j}+\Delta\psi_{k_{x},j}^{\dagger}(i\sigma_{y})\psi_{-k_{x},j}^{\dagger}+{\rm h.c}..
\end{multline}
Here, $2\pi\phi$ is the flux per unit cell, and $\varepsilon_{j}(k_{x},\phi)=-2t_{x}\mbox{cos}(k_{x}-2\pi\phi j)-\mu$.
We have used the Landau gauge, such that $A_{x}=By=2\pi\phi j$.

Applying the same procedure as in Sec. \ref{sec:spinfull model},
the Hamiltonian matrix at $k_{x}=0$ is given by:

\begin{multline}
h_{j}(k_{x}=0,\phi)=\varepsilon_{j}(0,\phi)\tau_{z}+V_{z}\sigma_{z}+\alpha\mbox{sin}(2\pi\phi j)\sigma_{y}+\Delta\tau_{x}.
\end{multline}
For a sufficiently small orbital field such that $\alpha eBL_{y}\ll\Delta,V_{z}$%
\footnote{Taking $\alpha=\Delta_{{\rm SO}}\lambda_{{\rm SO}}/\hbar$ where $\lambda_{{\rm SO}}$
is the a typical spin orbit coupling length and $\Delta_{{\rm SO}}$
is a typical spin orbit energy. Since $V_{z}=\frac{gB\hbar e}{2m_{e}}$,
the factor \textbf{$B$ }in the inequality condition cancels and we
find the condition $\Delta_{{\rm SO}}\ll\frac{g\hbar^{2}}{2m_{e}\lambda_{{\rm SO}}L_{y}}$.
For $\lambda_{{\rm SO}}=100nm$, $L_{y}=1\mu m$ the requirement is
$\Delta_{{\rm SO}}\ll0.01g[K]$. In the experiments \cite{das2012zero,mourik2012signatures}
this condition is only marginally satisfied. %
} (where $L_{y}$ is the width of the system), we can treat the orbital
field perturbatively. To zeroth order in $\phi$, the single wire
Hamiltonian at the critical point {[}Eq.(\ref{eq:critical}){]} is
diagonalized by a Bogoliubov transformation $W_{k_{x}}$ specified
in Eq.(\ref{eq:W}). In terms of the eigenstates $\psi_{1,k_{x},j}$
and $\psi_{2,k_{x},j}$, the vector potential term is

\begin{multline}
-\alpha2\pi\phi j\psi_{k_{x},j}^{\dagger}\sigma_{y}\psi_{k_{x},j}+\mathrm{h.c.}\\
=i\alpha\frac{\Delta^{2}(1+s^{2})}{(V_{z}s)^{2}}2\pi\phi j\psi_{2,k_{x},j}^{\dagger}\psi_{1,k_{x},j}^{\dagger}+{\rm h.c}..
\end{multline}
Here, $s\equiv\frac{-2t_{x}-\mu+V_{z}}{\Delta}$. Upon projecting
the Hamiltonian onto the low-energy ($\psi_{1,k_{x},j}$) sector,
this term is zero as it couples $\psi_{1,k_{x},j}$ and $\psi_{2,k_{x},j}$.
Therefore, to first order in $\phi$, the inter-wire Hamiltonian projected
to the low-energy subspace retains the form of Eq. (\ref{eq:Kitaev's H spinfull}).
I.e., to first order in $\phi$, the ``sweet point'' condition {[}Eq.(\ref{eq:"sweat point"-1}){]}
in unaffected.

\bibliographystyle{apsrev4-1}
\bibliography{bibliography}

\end{document}